%% file: ms.tex
\documentclass[12pt,preprint]{aastex}
\shorttitle{Seven southern hemisphere VLBI pulsar parallaxes}
\shortauthors{Deller et al.}

\begin{document}

\newcommand{\pbdot}{\ensuremath{\dot{P}_{\mathrm b}}}
\newcommand{\degrees}{\ensuremath{\,^{\circ}}}
\newcommand{\pone}{PSR J0108--1431}
\newcommand{\ptwo}{PSR J0437--4715}
\newcommand{\pthree}{PSR J0630--2834}
\newcommand{\pfour}{PSR J0737--3039A/B}
\newcommand{\pfive}{PSR J1559--4438}
\newcommand{\psix}{PSR J2048--1616}
\newcommand{\pseven}{PSR J2144--3933}
\newcommand{\peight}{PSR J2145--0750}
\newcommand{\kms}{\ensuremath{\mathrm {km\ s}^{-1}}}

\title{Precision southern hemisphere VLBI pulsar astrometry II: measurement of seven parallaxes}

\author{A. T. Deller,\altaffilmark{1} S. J. Tingay,\altaffilmark{2}, M. Bailes,\altaffilmark{1}, 
J. E. Reynolds\altaffilmark{3}}

\altaffiltext{1}{Centre for Astrophysics and Supercomputing, Swinburne University of Technology, Mail H39, P.O. Box 218, Hawthorn, VIC 3122, Australia}

\altaffiltext{2}{Department of Imaging and Applied Physics, Curtin University of Technology, Bentley, WA, Australia}

\altaffiltext{3}{Australia Telescope National Facility, Epping, NSW, Australia}

\begin{abstract}
Accurate measurement of pulsar distances via astrometry using very long baseline interferometry
enables the improvement of Galactic electron density distribution models, improving distance estimates 
for the vast majority of pulsars for which parallax measurements are unavailable. However,
pulsars at southern declinations have been under--represented in previous interferometric astrometry 
campaigns, due to the dominance of northern hemisphere instrumentation for
astrometry.  In order to redress this imbalance, we have conducted a two--year astrometric 
campaign targeting eight southern pulsars with the Australian Long Baseline Array. 
The program summarized in this paper has resulted in the measurement of seven new pulsar 
parallaxes, with success on objects down to a mean flux density of 800 $\mu$Jy at 1600 MHz.
Our results highlight the substantial uncertainties that remain when utilizing free electron density 
models for individual pulsar distances.
Until this study, PSR J0630--2834 was believed to convert 16\% of its spin--down
energy into x--rays, but our measured parallax distance of $332^{+52}_{-40}$\, pc 
has revised this value to $<$1\%.  In 
contrast, PSR J0108--1431 was found to be almost a factor of two more distant than previously thought,
making its conversion of spin--down energy to x--rays the most efficient known ($>$1\%).
The 8.5 second radio pulsar J2144--3933 was found to be closer than previously predicted, 
making its apparent 1400 MHz radio luminosity the lowest of any known pulsar (20 
$\mu$Jy kpc$^{2}$).  We have examined the growing population of neutron stars with accurate 
parallaxes to determine the effect of distance errors on the underlying neutron star velocity
distribution, and find that typical distance errors may be biasing the estimated mean pulsar velocity 
upwards by 5\%, and are likely to exaggerate the distribution's high--velocity tail.
\end{abstract}


\section{Introduction}
Studies of pulsars are often plagued by the uncertainty in distance--dependent quantities 
introduced by the reliance on dispersion measure (DM) distance estimates.  Due to the 
non--uniform distribution of ionized material in the ISM on sub-kpc scales \citep{cordes07a}, 
the correspondence between a pulsar's DM and distance is often uncertain.  
The two most widely used Galactic electron density distribution models are those presented by
\citet{taylor93a}, hereafter referred to as TC93, and \citet{cordes02a}, hereafter referred to
as NE2001.  Although distances estimated from these models are typically quoted as being accurate
to 20\%, previous astrometric pulsar observations have shown that much greater errors are
possible for individual objects 
\citep[e.g. a factor of 2.5 error in the distance for PSR~B0656+14;][]{brisken03a}.

Independent distance measures to pulsars are extremely valuable, both to enable the confident 
estimation of distance--dependent parameters for individual pulsars and to refine
DM--based distance models for the remainder of the pulsar population. 
They can be provided from HI absorption measurements 
\citep[e.g.][]{koribalski95a}, association with other astrophysical objects \citep[e.g.][]{camilo06a}, 
and annual geometric parallax measurements made via timing \citep[e.g.][]{hotan06a} or 
VLBI \citep[e.g.][]{brisken02a}.  Both scientific justifications
are more acute  at southern declinations, where prior to this observing program 
only two published VLBI pulsar parallaxes were available \citep{bailes90a,dodson03a}, despite the
fact that 70\% of known pulsars are located at a declination $<0$\degrees.  Recently, a further
four parallaxes have been published for pulsars with declinations between 0\degrees\ and
$-20$\degrees\ \citep{chatterjee09a}.  At the present time, however, Galactic electron models still have 
fewer constraints in the south than at northern declinations, 
and even the most scientifically desirable targets such as the double pulsar J0737--3039A/B 
generally have only a DM distance for deriving luminosities, 
velocities, population sizes and so on.

The eight pulsars targeted by this survey are described in Table~\ref{tab:pulsars}.  Two of these
(\pfive\ and \psix) were chosen simply as relatively bright nearby pulsars on 
which our data reduction techniques could be tested in the high signal--to--noise regime.  The 
remainder were each chosen because an independent distance measure offered
a resolution to an outstanding question surrounding the pulsar.  
Further constraints on target selection 
were imposed by the need for relatively uniform distribution around a 
range of right ascensions, due to observational logistics.  Thus, the target selection in
no way comprises an unbiased sample of southern hemisphere pulsars; however, as noted
above, the paucity of existing parallax measurements in the south means that any
additions will be extremely useful in constraining electron density distribution models at these
declinations.

The six pulsars selected on scientific merit can be
broadly grouped into two categories.  The first encompasses those possessing unusual 
luminosity characteristics (low radio luminosity pulsars \pone\ and \pseven, and the high x--ray 
luminosity pulsar \pthree) where the primary motivation is to obtain a confident estimate of the
pulsar luminosity.  The second group encompasses binary
pulsars used for tests of general relativity (GR; \ptwo, \pfour\ and \peight), where accurate distances
and velocities are required to calculate kinematic contributions to pulsar timing terms
\citep[the Shklovskii effect;][]{shklovskii70a}.  

In this paper, we summarize our observational strategy in \S\ref{sec:obs}, and present the results of
the survey in \S\ref{sec:results}.  Interpretation of the revised distance measure for each pulsar is
presented in \S\ref{sec:discussion}.  While the results for \ptwo, \pfour, and \pfive\ have been
presented previously \citep{deller08b,deller09a,deller08a}, a summary is included here for
completeness.  The accuracy of TC93 and NE2001 predictions for pulsar distances are
examined in \S\ref{sec:discussion}, along with an investigation into the likely impacts of distance model
errors on derived pulsar population parameters.  Our conclusions are presented in 
\S\ref{sec:conclusions}.

\section{Observations}
\label{sec:obs}
The observational methodology employed has  been discussed in detail in \citet{deller08a}, and 
a brief overview is presented here.  Seven observing sessions\footnote{As noted in 
\citet{deller08a}, it was necessary to discard the first of eight observed epochs (MJD 53870) 
due to the probability of uncorrectable ionospheric model errors} spanned an
18 month period between August 2006 and February 2008, with a subset of the eight pulsars 
observed at each epoch, depending on which were closest to parallax extrema. 
All observations were made at 1650 MHz, with the exception of PSR J0437--4715 which was
observed separately at 8400 MHz due to its brightness and narrow pulse profile.
A phase reference cycle time of six minutes, allocated equally between calibrator and
target, was used for all observations.

Where possible, all six stations of the Australian Long Baseline Array (LBA) were used for 
observing.  However, the Ceduna telescope does not possess a 1600 MHz receiver, and the 
use of a NASA DSN antenna\footnote{The 70m and 34m antennas were used in different
epochs} in Tidbinbilla was possible only when 
telescope time was available, which was for roughly half of the observational program.  
Figure~1 of \citet{deller08a} shows representative $uv$ coverage for 1650 MHz and 8400 
MHz observations.  While higher data rates were recorded at some stations, the final astrometric
processing of the observations utilized four single polarization 16 MHz bands sampled at 
two bit precision, yielding a total data rate of 256 Mbps per telescope.

\begin{figure}
\begin{center}
\begin{tabular}{cc}
\includegraphics[width=0.45\textwidth, angle=270, clip]{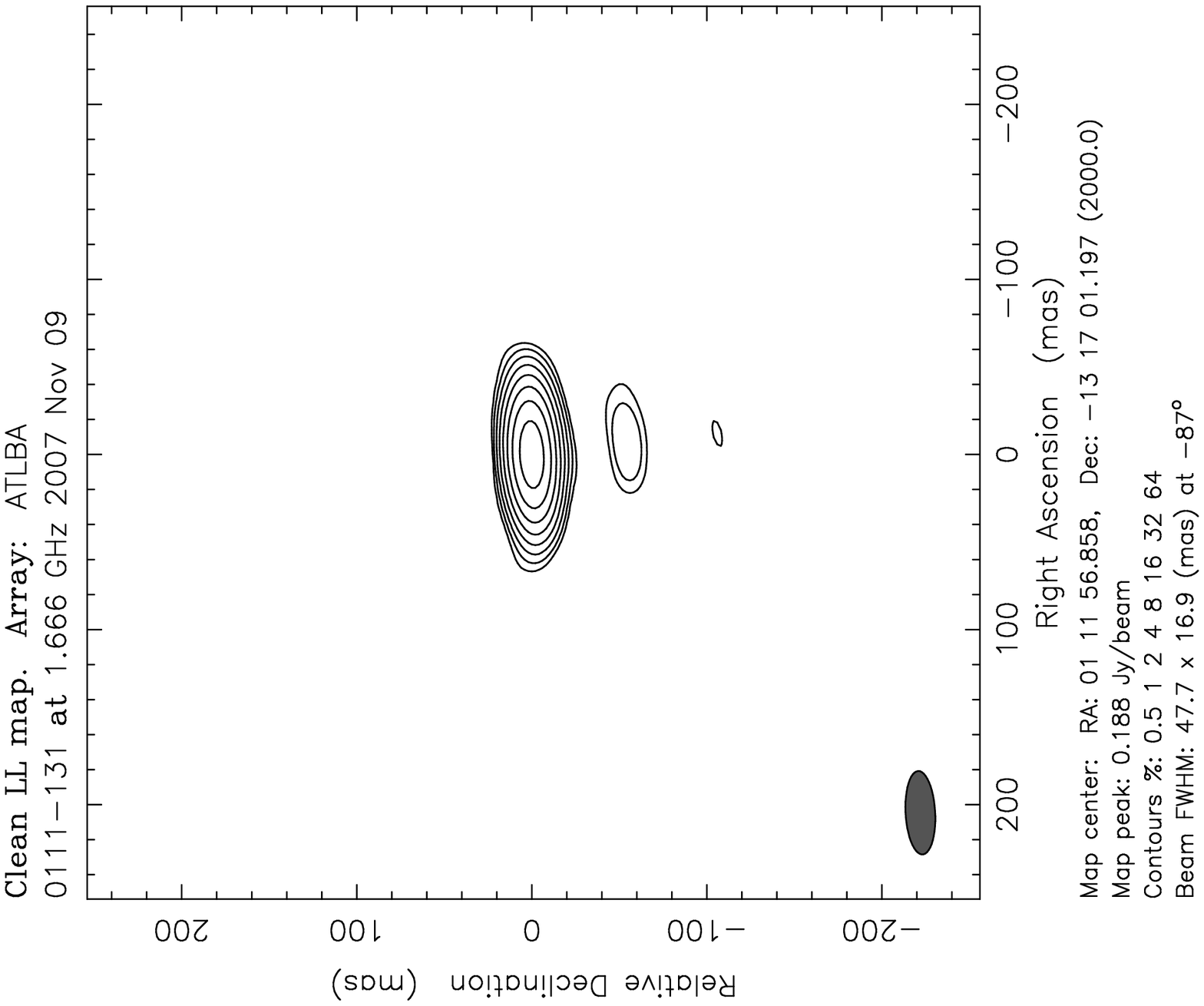} &
\includegraphics[width=0.45\textwidth, angle=270, clip]{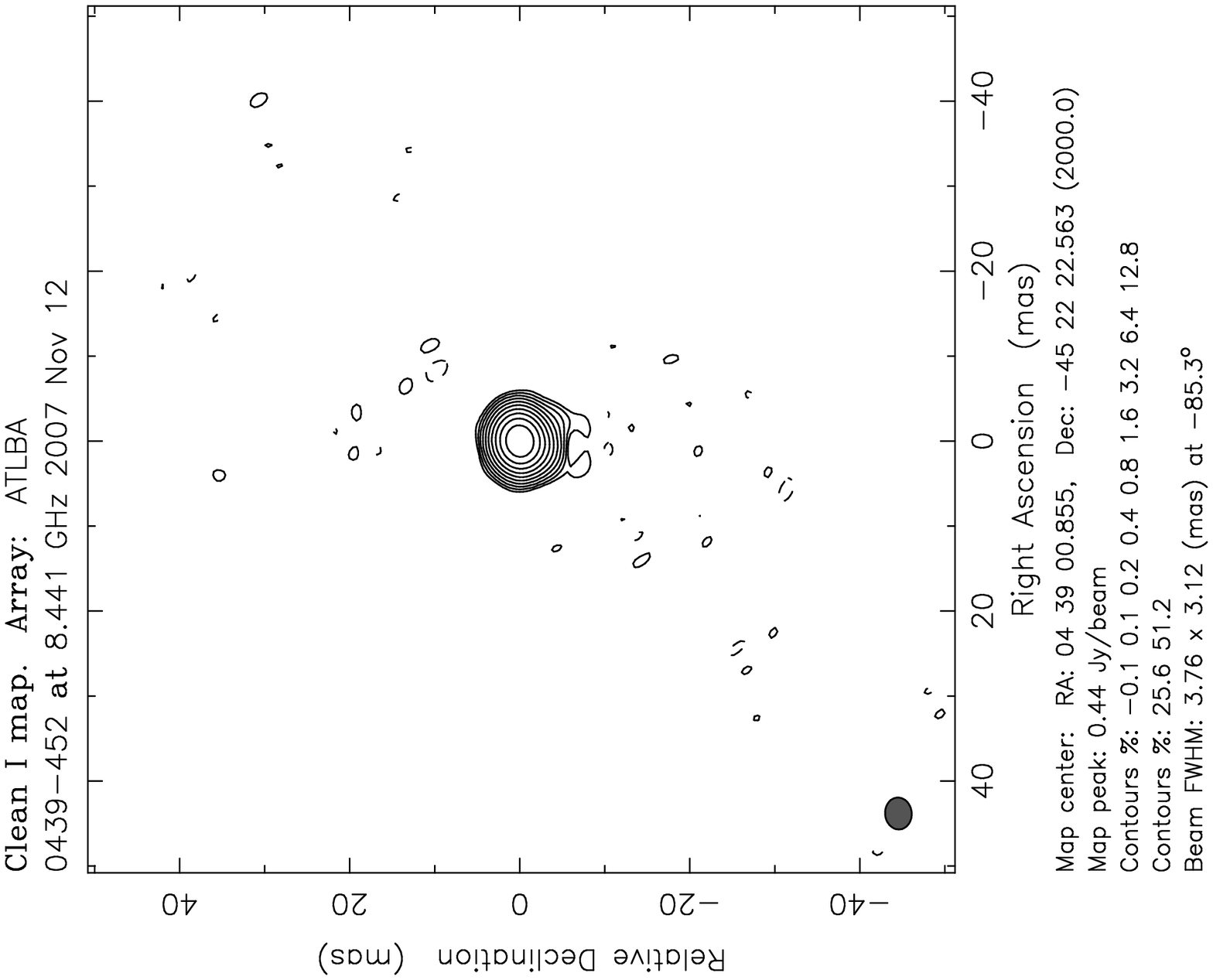} \\
\includegraphics[width=0.45\textwidth, angle=270, clip]{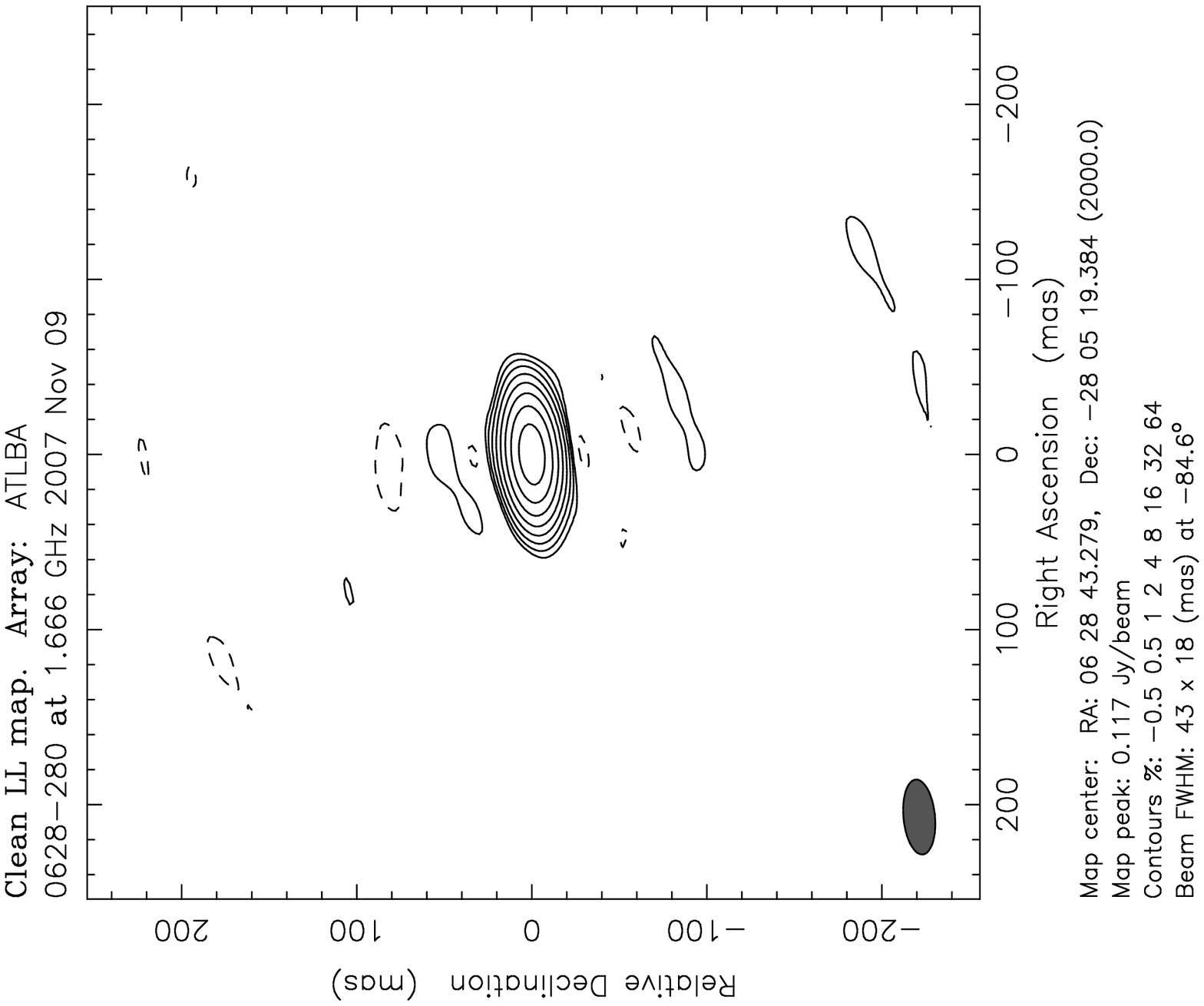} &
\includegraphics[width=0.45\textwidth, angle=270, clip]{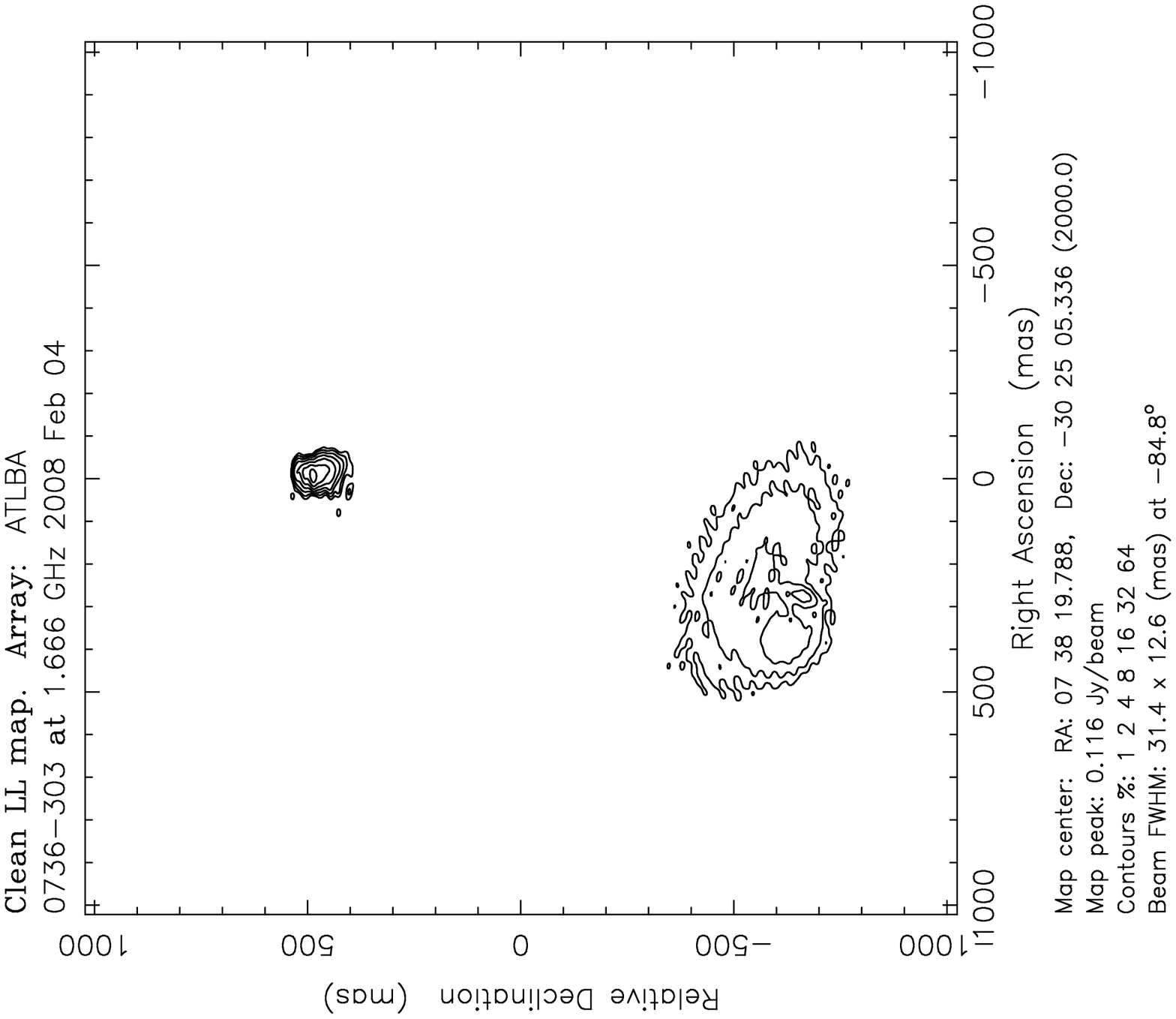} \\
\end{tabular}
\caption{Images of the phase reference sources used.  In all instances, the image contours increase
by factors of two from a minimum value expressed as a percentage of peak image flux.  Runnning
left to right, top to bottom, the calibrator name, peak image flux (mJy) and minimum contour are:
J0111--1431, 188, 0.5\%; J0439--4522, 440, 0.1\%; J0628--2805, 117, 0.5\%; B0736--303, 116, 1\%.
The image of B0736--303 has been shifted to place the image centre between the two components
for clarity.}
\label{fig:cals1}
\end{center}
\end{figure}

\begin{figure}
\begin{center}
\begin{tabular}{cc}
\includegraphics[width=0.45\textwidth, angle=270, clip]{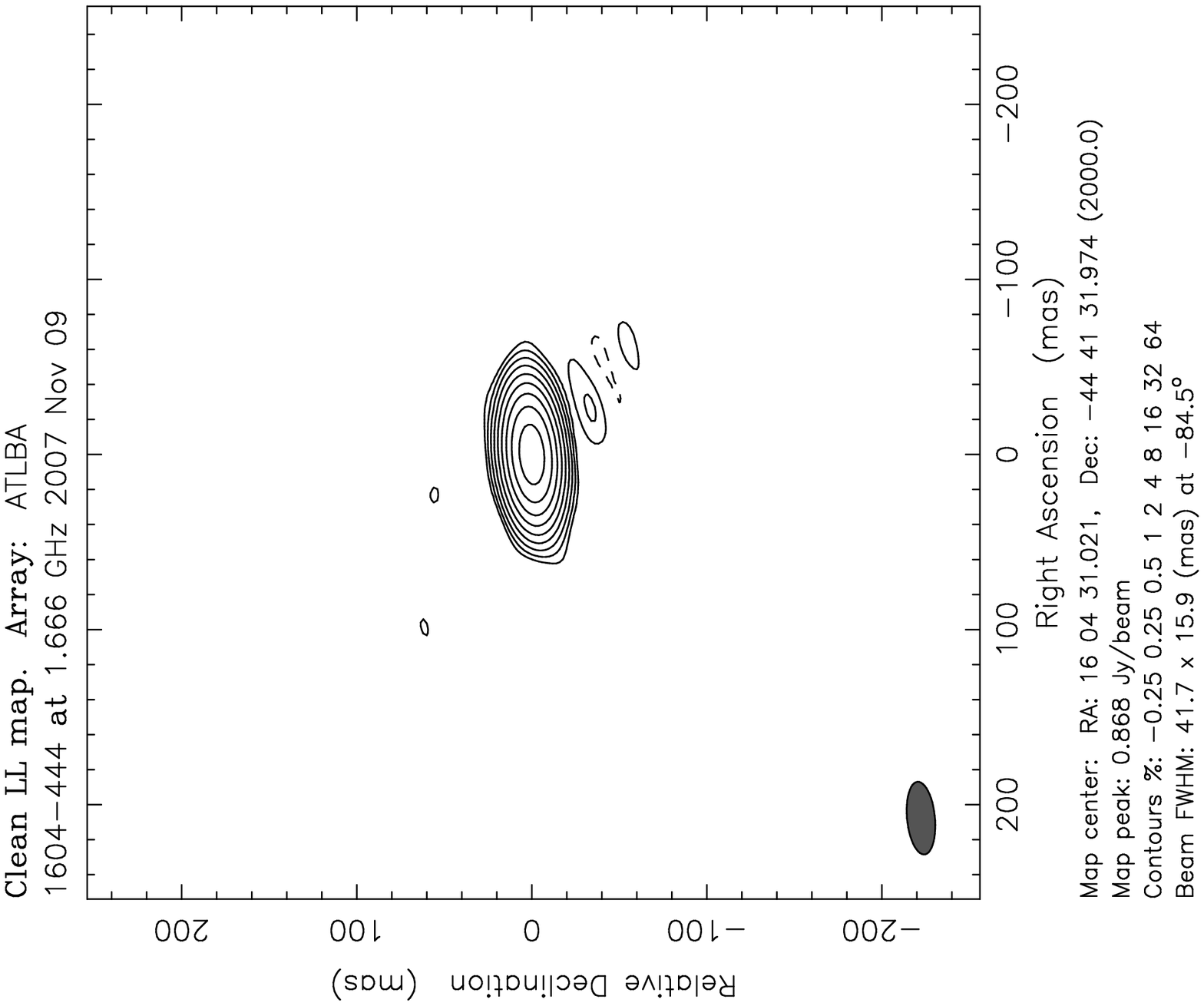} &
\includegraphics[width=0.45\textwidth, angle=270, clip]{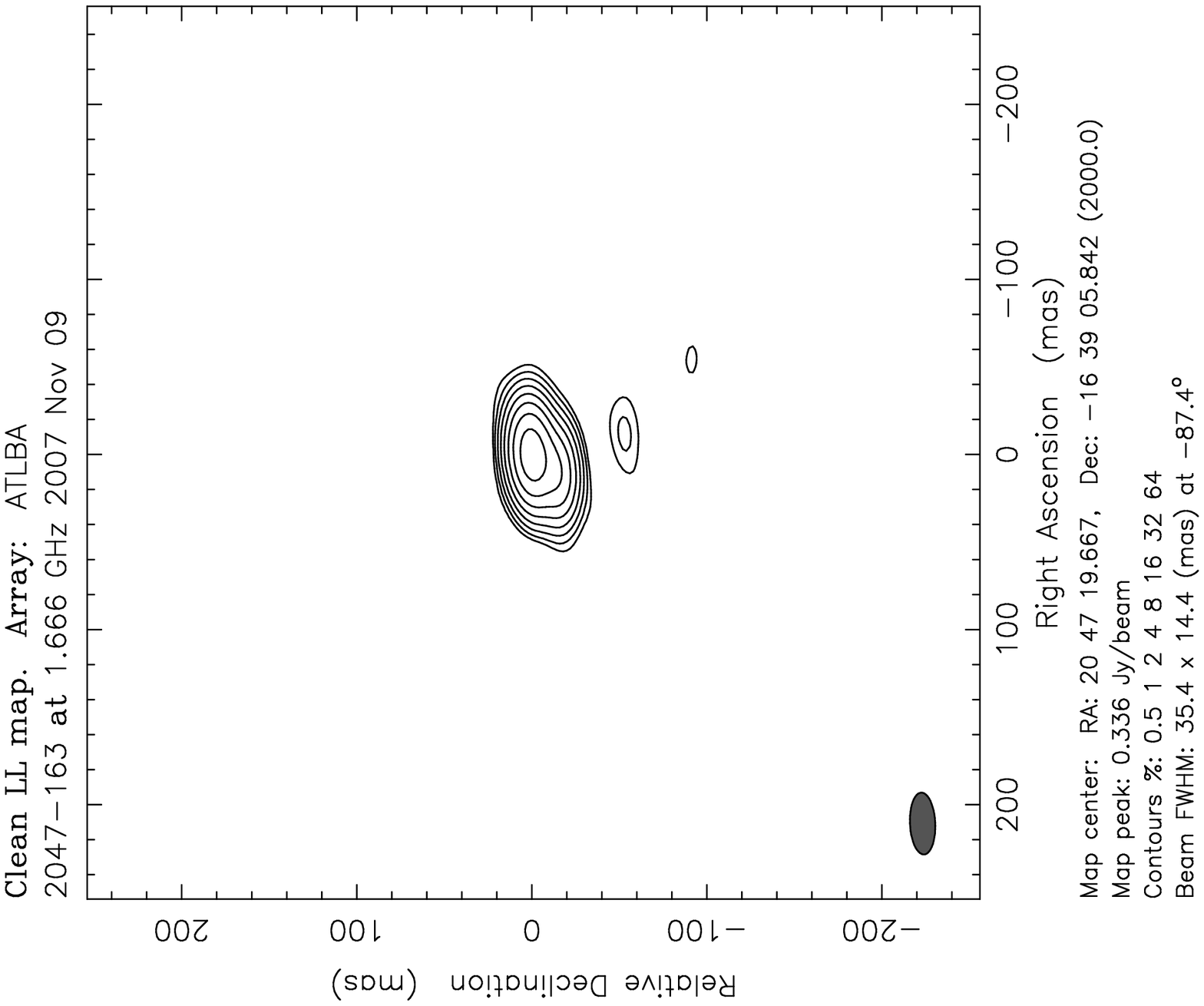} \\
\includegraphics[width=0.45\textwidth, angle=270, clip]{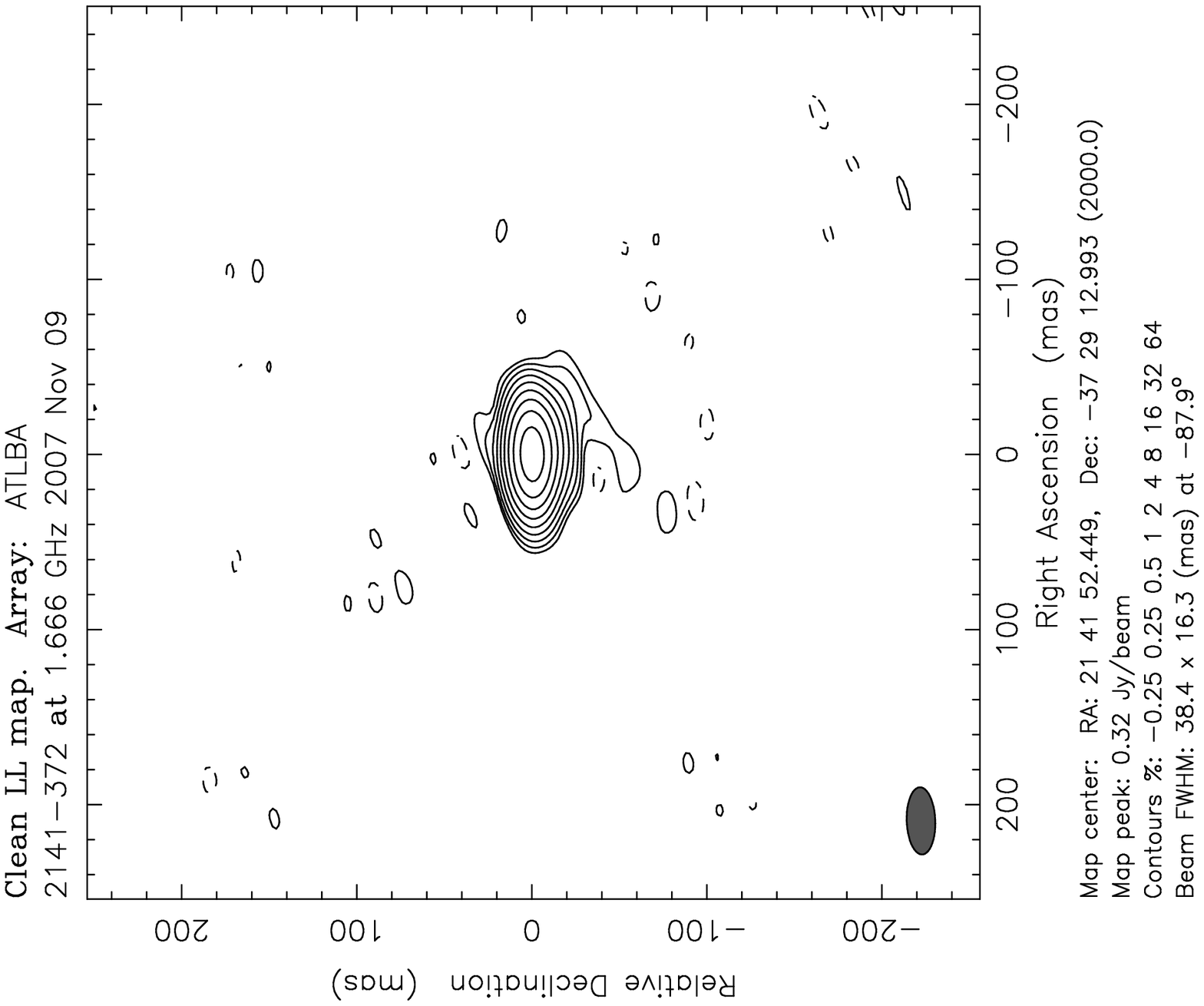} &
\includegraphics[width=0.45\textwidth, angle=270, clip]{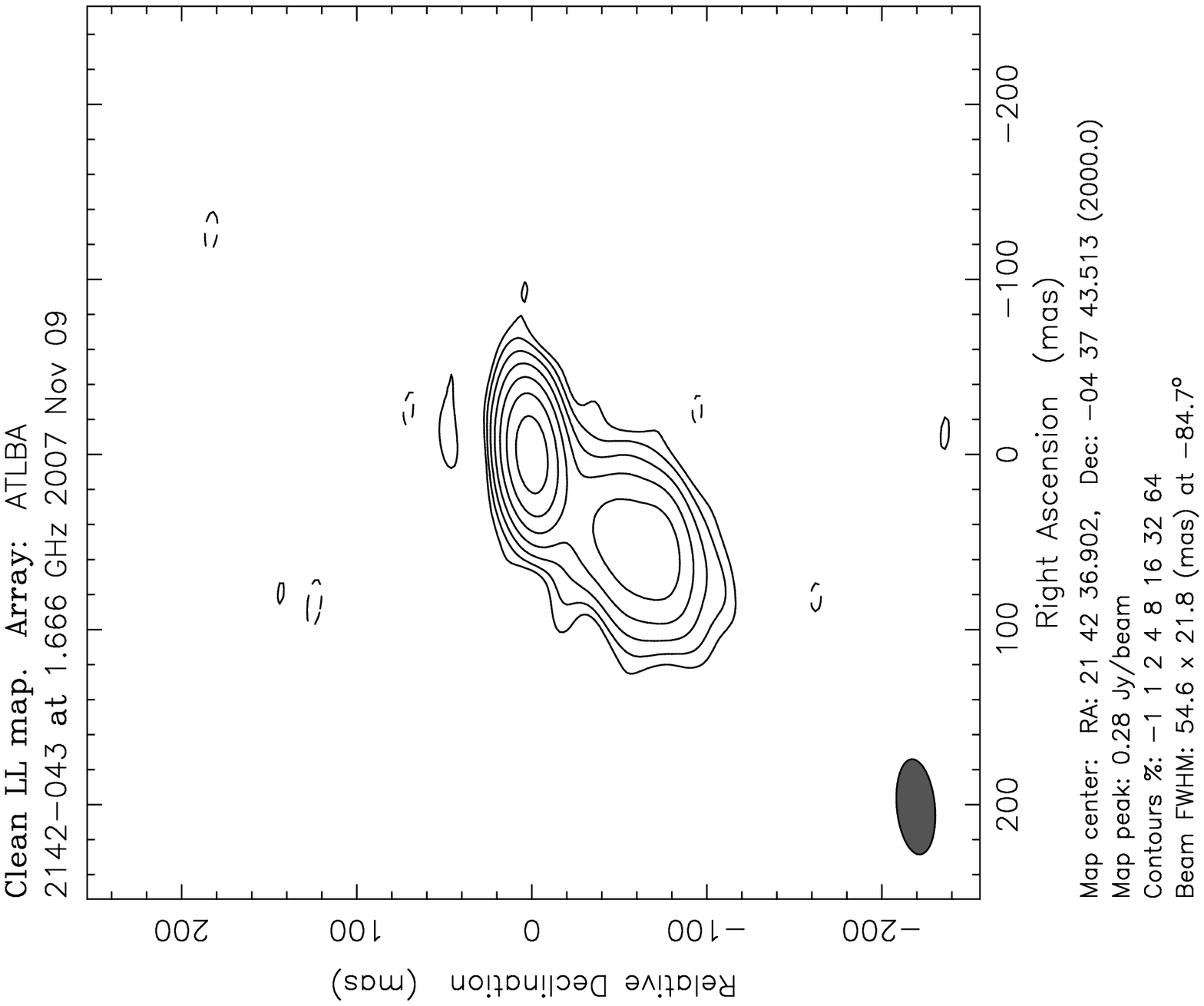} \\
\end{tabular}
\caption{Images of the phase reference sources used.  In all instances, the image contours increase
by factors of two from a minimum value expressed as a percentage of peak image flux.  Runnning
left to right, top to bottom, the calibrator name, peak image flux (mJy) and minimum contour are:
J1604--4441, 868, 0.25\%; J2047--1639, 336, 0.5\%; J2141--3729, 320, 0.25\%; J2142--0437, 280, 
1\%.}
\label{fig:cals2}
\end{center}
\end{figure}

The data were correlated using matched filtering (a more sophisticated pulsar `gate') 
on pulse profiles with the DiFX software correlator \citep{deller07a}, which increases the
signal to noise ratio of the correlated dataset by weighting the visibility data according to the 
predicted pulsar flux. The resultant
correlated datasets were processed with a pipeline utilising the Parseltongue interface
\citep{kettenis06a} to AIPS\footnote{http://www.aoc.nrao.edu/aips}.
This pipeline, which is described in detail in \citet{deller08a}, covered calibration, flagging, 
geometric and ionospheric model correction, and fringe--fitting, and corrected for
reference source structure and pulsar scintillation.  Figures~\ref{fig:cals1} and \ref{fig:cals2} show
the composite reference images used for the each phase reference source, and the position centroid
used for each calibrator is given in Table~\ref{tab:calpos}.  Station positions were taken from the global
VLBI solution 2008b\_astro\footnote{http://lacerta.gsfc.nasa.gov/vlbi/solutions/2008b\_astro/}, which includes
new position determinations for the Ceduna, Mopra and ATCA stations as described in \citet{petrov09a}.
The calibrated pulsar images produced at each epoch were then fit for position, 
and the multi--epoch position dataset thus generated
was fit for reference position, proper motion and parallax.  The astrometric fit at the final stage
of the pipeline estimated and attempted to account for the presence of systematic errors in
the pulsar position measurements.

\section{Results}
\label{sec:results}
Of the eight pulsars observed, parallaxes were obtained for seven.  The remaining pulsar, 
\peight, appeared to suffer heavily from refractive scintillation, as was reported during a previous 
unsuccessful attempt at VLBI astrometry using the VLBA \citep{brisken02a}.  In our
observations, it was detected in only two usable epochs,  and so while a proper motion can
be derived by holding the parallax value fixed over a reasonable range of values, 
the results should be treated with some caution.

Table~\ref{tab:allresults} summarizes the results for the eight target pulsars. All errors are 1$\sigma$\ 
values obtained using the ``inclusive" fit approach described in \citet{deller08a},
which models the presence of systematic errors.  The resultant reduced chi--squared value is 1.0 for all
pulsars except \pfour\ (for which the initial fit gave a reduced chi--squared value of 0.79, and hence
no additional systematic error component was required).  All positions are specified
at MJD 54100 (31 December 2006) and errors on positions are fit errors which do not include the 
several mas uncertainty typical to calibrator positions and constant phase referencing offsets.
Table~\ref{tab:allresults} lists the covariances between proper motion and parallax for each 
pulsar, and shows that the addition of future epochs could considerably improve the parallax
precision for \ptwo\ and \psix\ in particular.

\begin{figure}
\begin{center}
\begin{tabular}{c}
\includegraphics[width=0.45\textwidth]{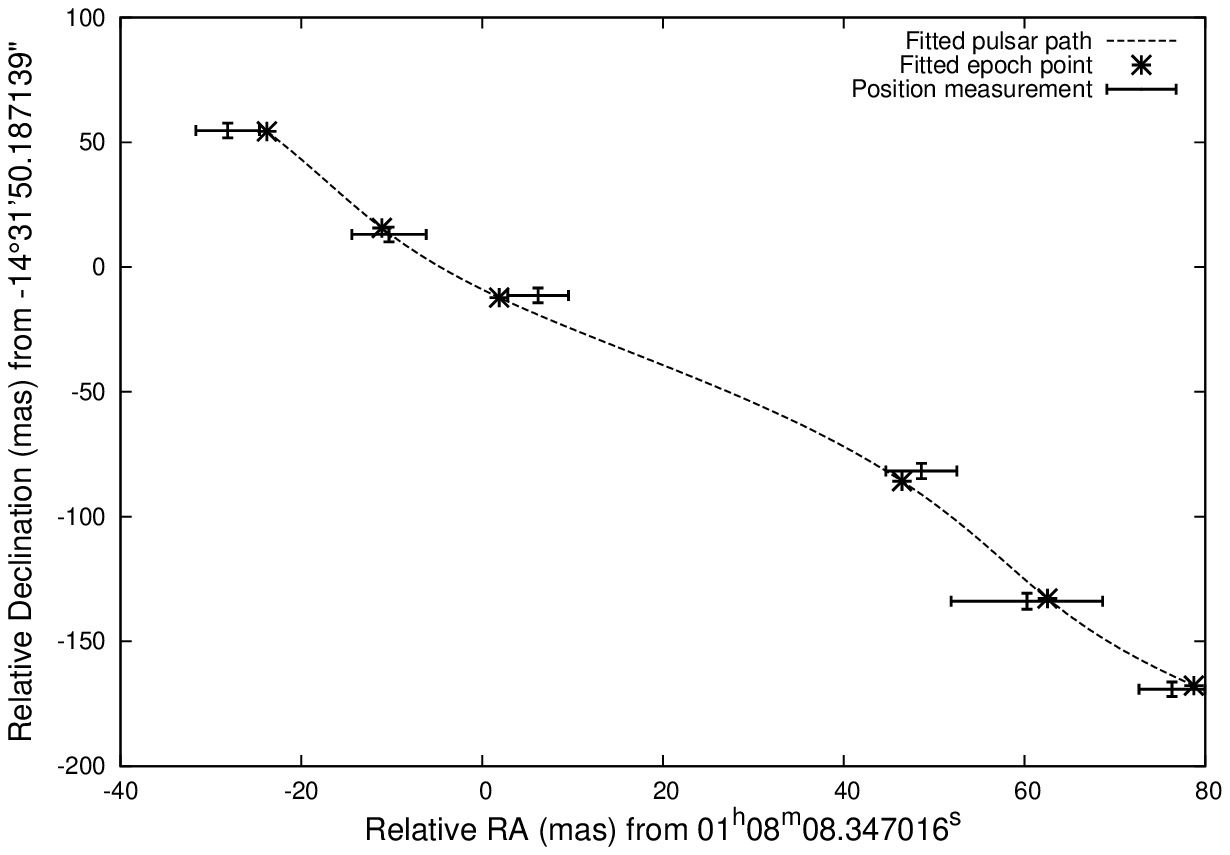} \\
\includegraphics[width=0.45\textwidth]{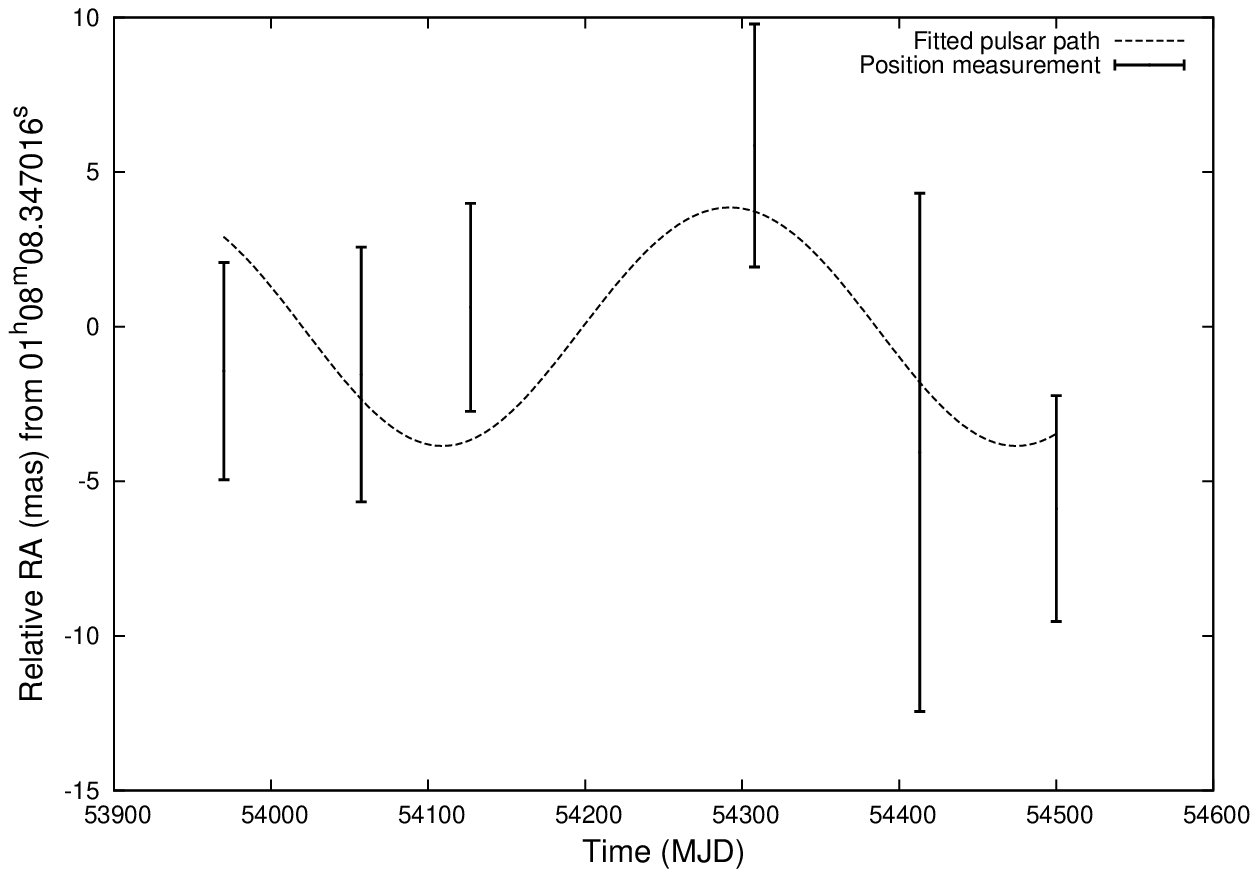} \\
\includegraphics[width=0.45\textwidth]{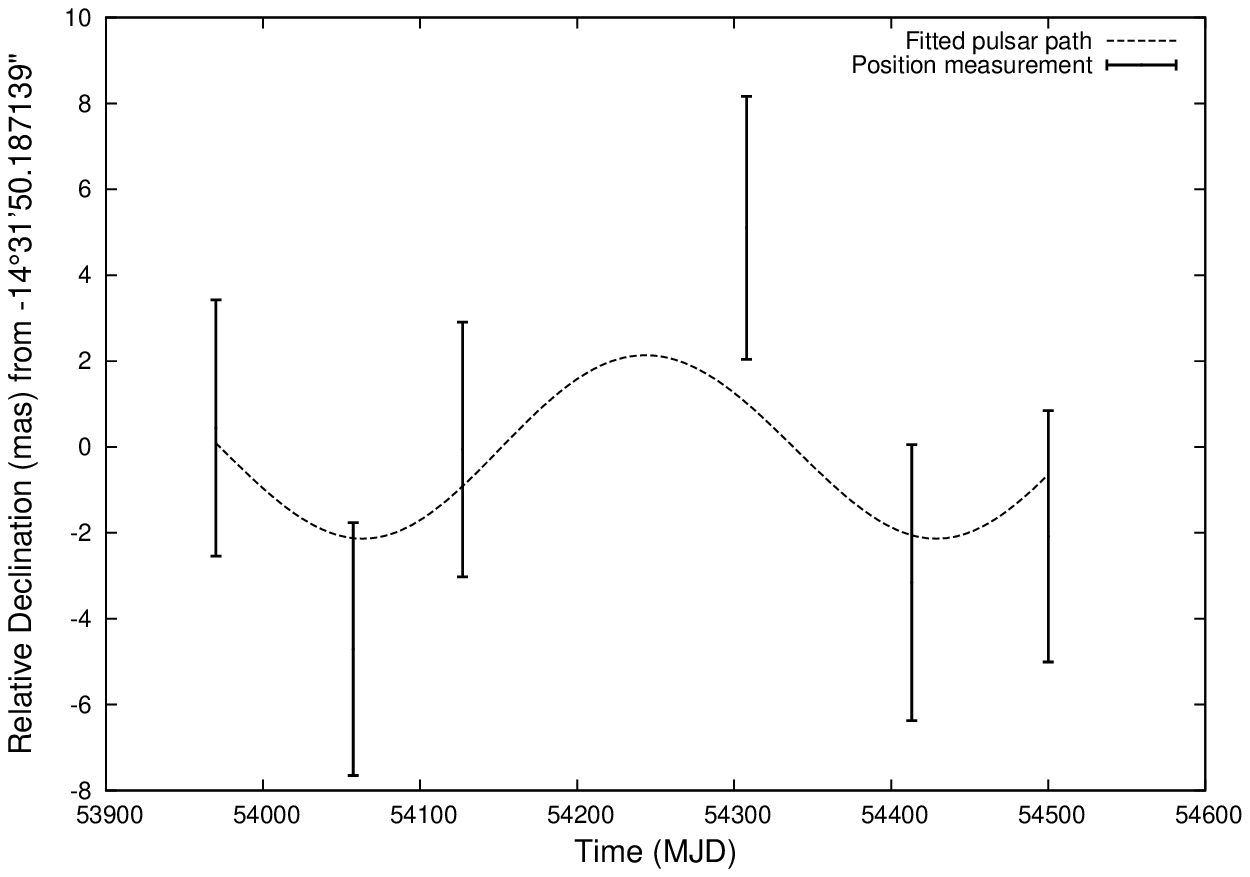} \\
\end{tabular}
\caption{Motion of \pone, with measured positions overlaid on the best fit. From top to bottom; 
motion in declination vs right ascension, motion in right ascension vs time with 
proper motion subtracted, and motion in declination vs time with proper motion subtracted.}
\label{fig:0108fit}
\end{center}
\end{figure}

\begin{figure}
\begin{center}
\begin{tabular}{c}
\includegraphics[width=0.45\textwidth]{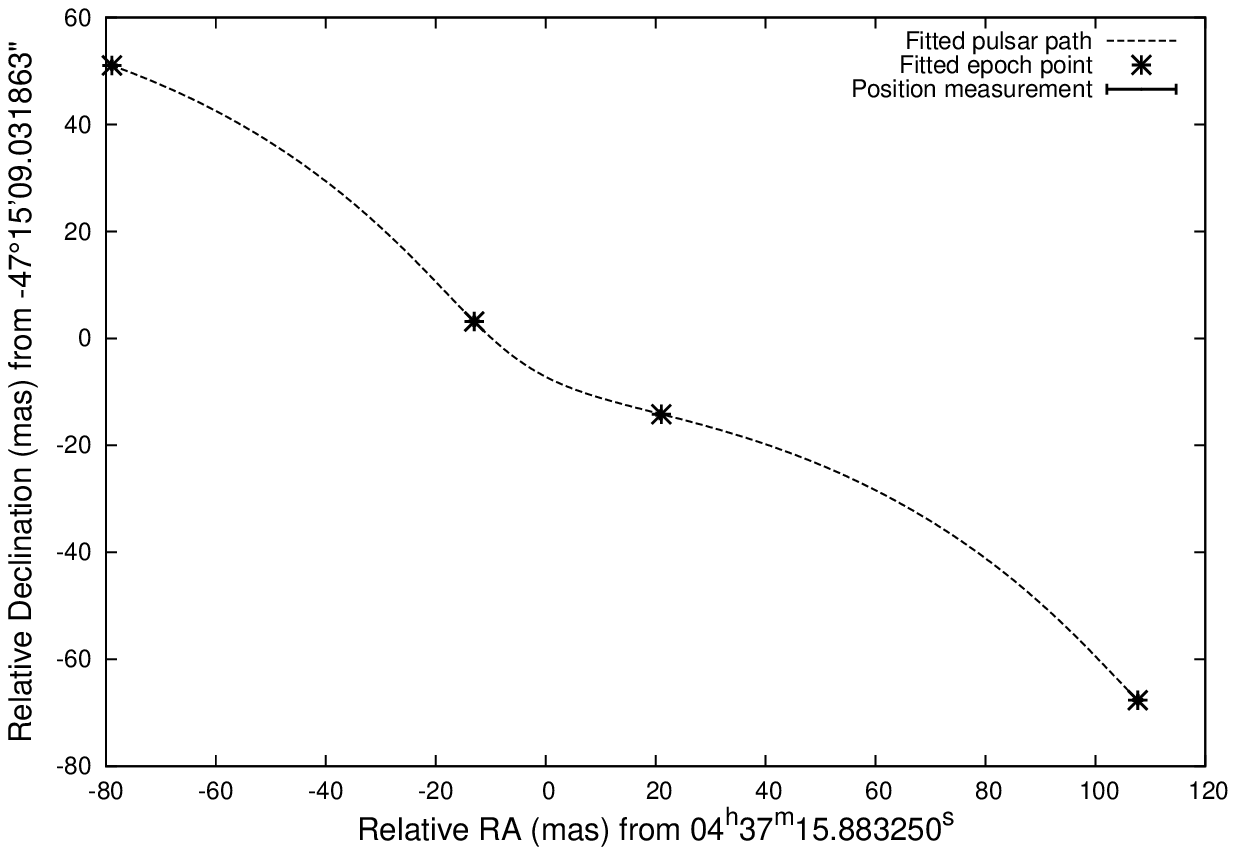} \\
\includegraphics[width=0.45\textwidth]{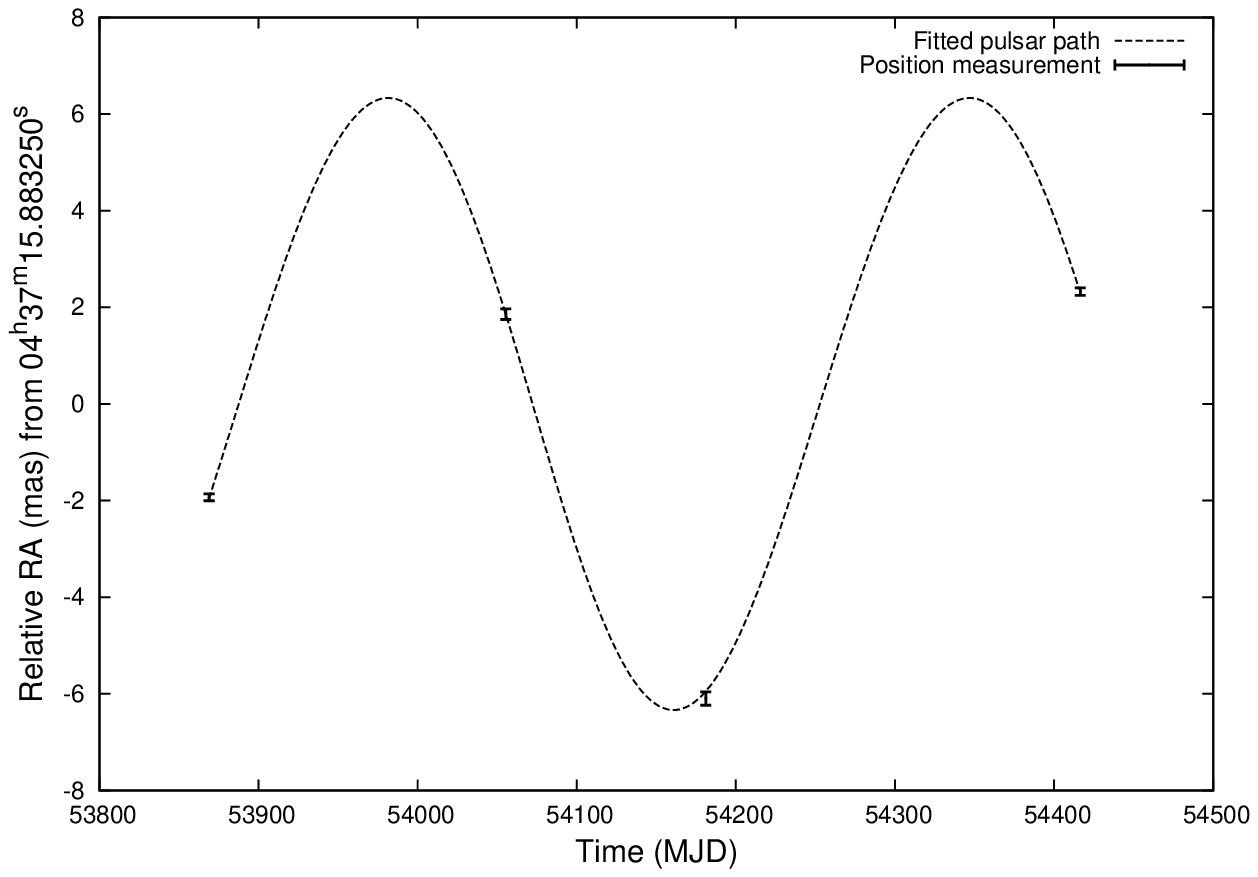} \\
\includegraphics[width=0.45\textwidth]{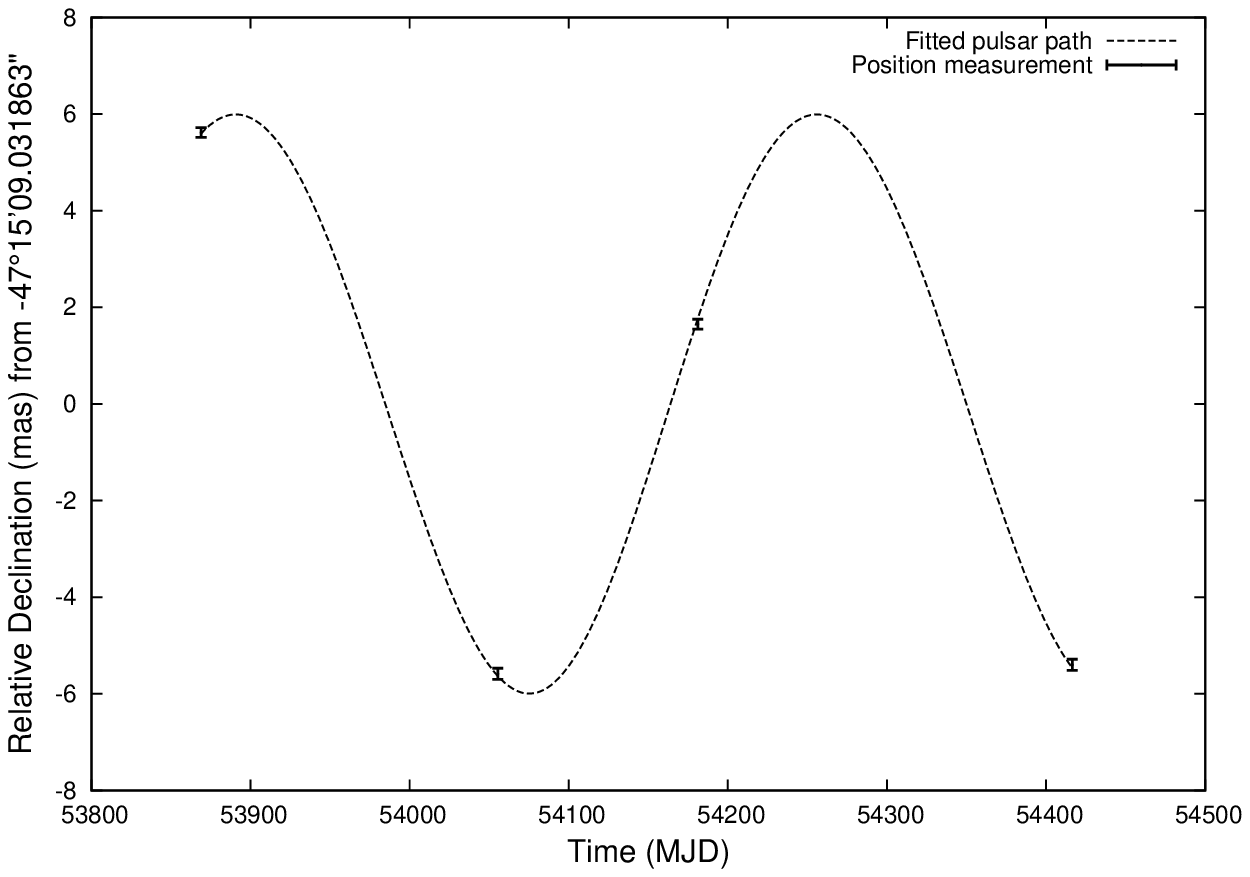} \\
\end{tabular}
\caption{Motion of \ptwo, with measured positions overlaid on the best fit. From top to bottom; 
motion in declination vs right ascension, motion in right ascension vs time with 
proper motion subtracted, and motion in declination vs time with proper motion subtracted.}
\label{fig:0437fit}
\end{center}
\end{figure}

\begin{figure}
\begin{center}
\begin{tabular}{c}
\includegraphics[width=0.45\textwidth]{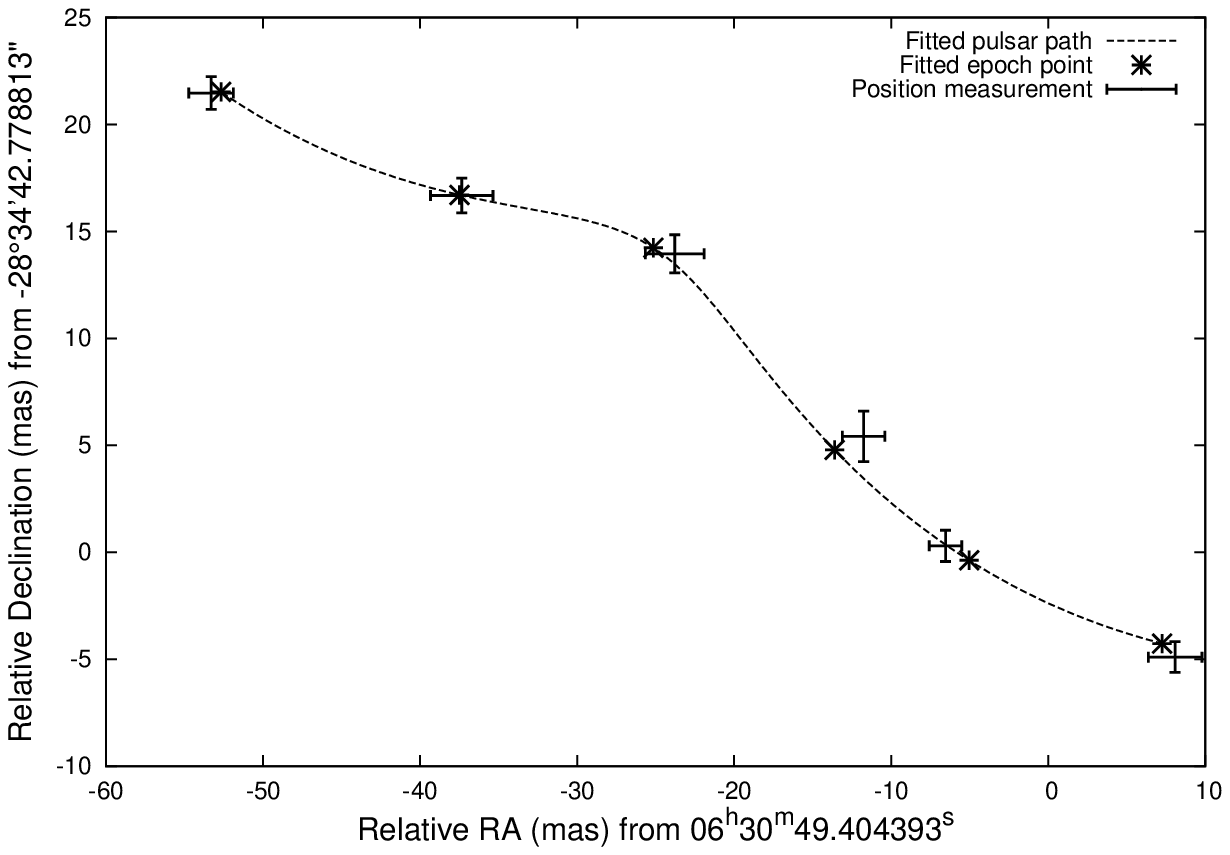} \\
\includegraphics[width=0.45\textwidth]{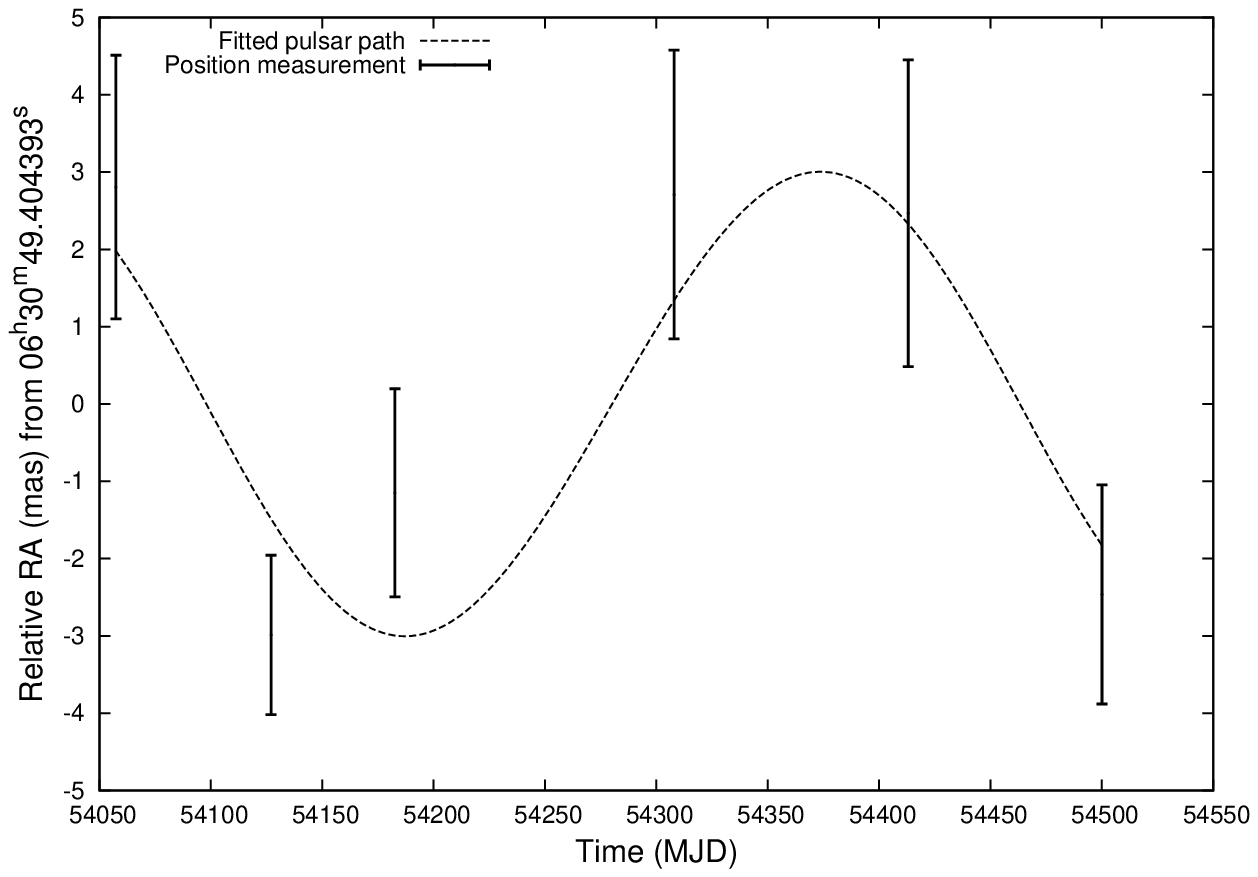} \\
\includegraphics[width=0.45\textwidth]{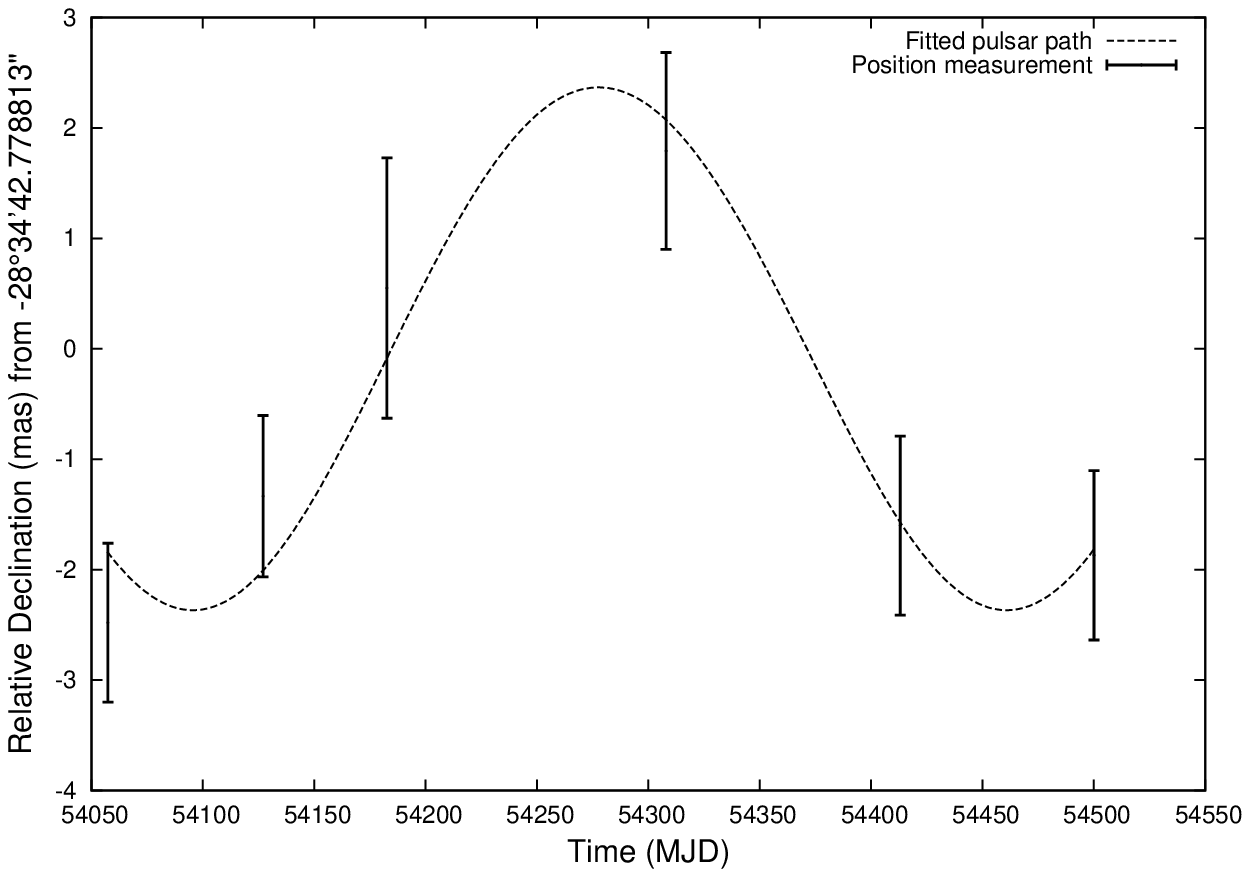} \\
\end{tabular}
\caption{Motion of \pthree, with measured positions overlaid on the best fit. From top to bottom; 
motion in declination vs right ascension, motion in right ascension vs time with 
proper motion subtracted, and motion in declination vs time with proper motion subtracted.}
\label{fig:0630fit}
\end{center}
\end{figure}

\begin{figure}
\begin{center}
\begin{tabular}{c}
\includegraphics[width=0.45\textwidth]{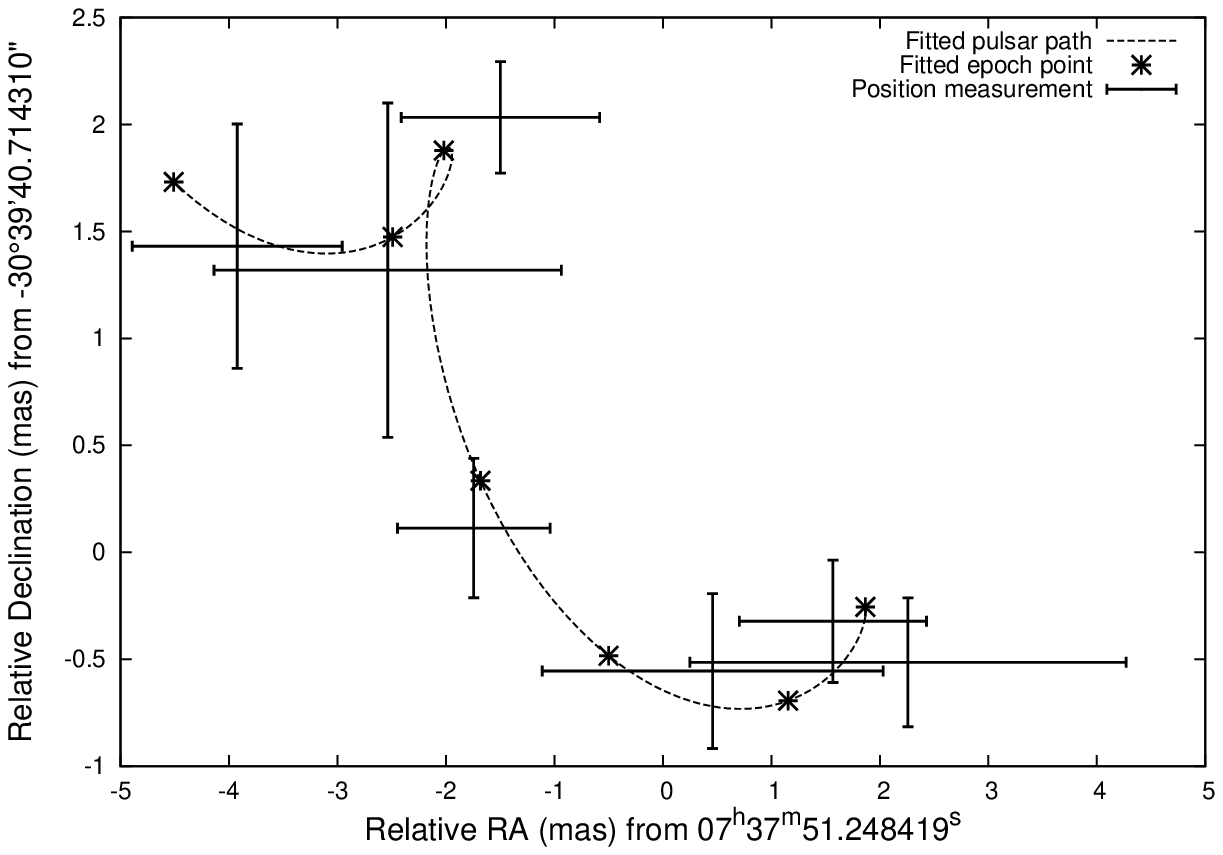} \\
\includegraphics[width=0.45\textwidth]{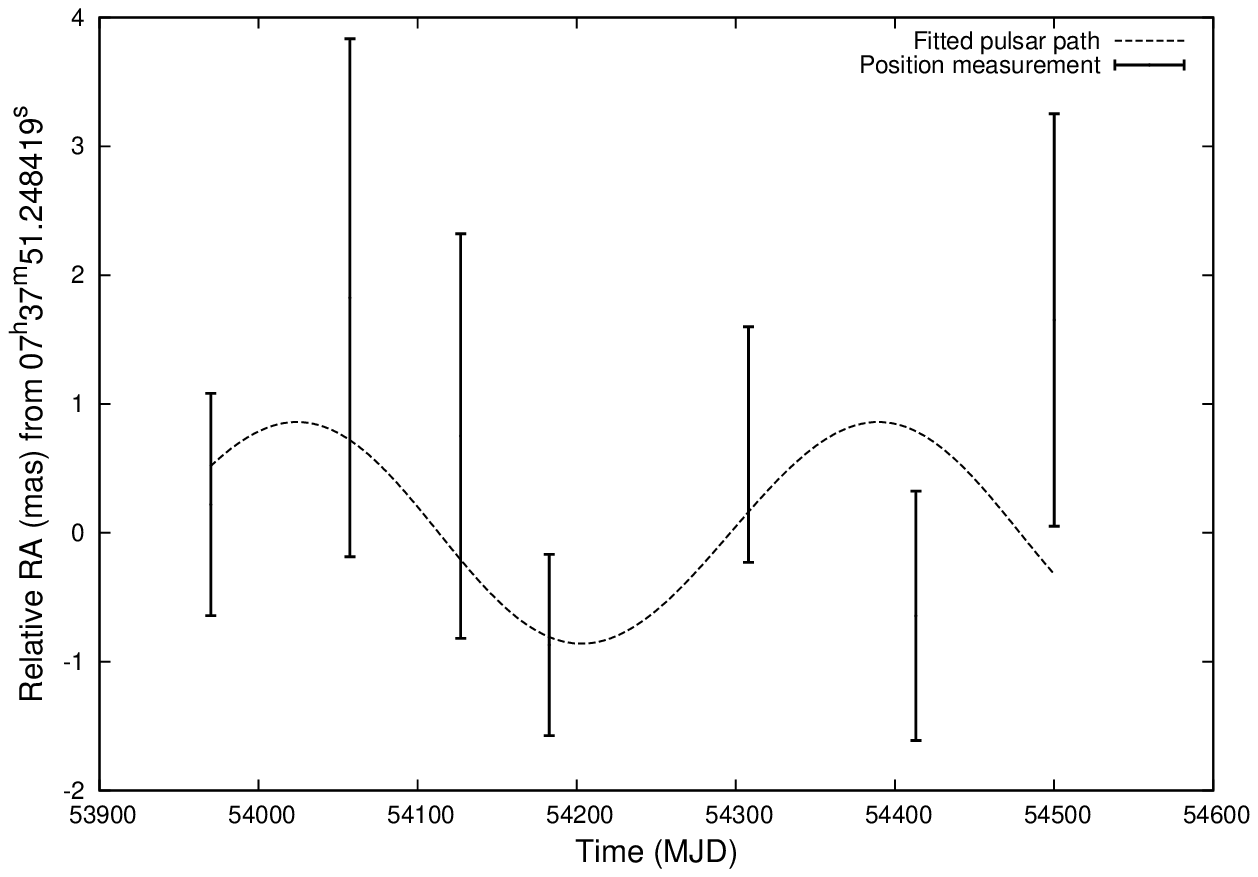} \\
\includegraphics[width=0.45\textwidth]{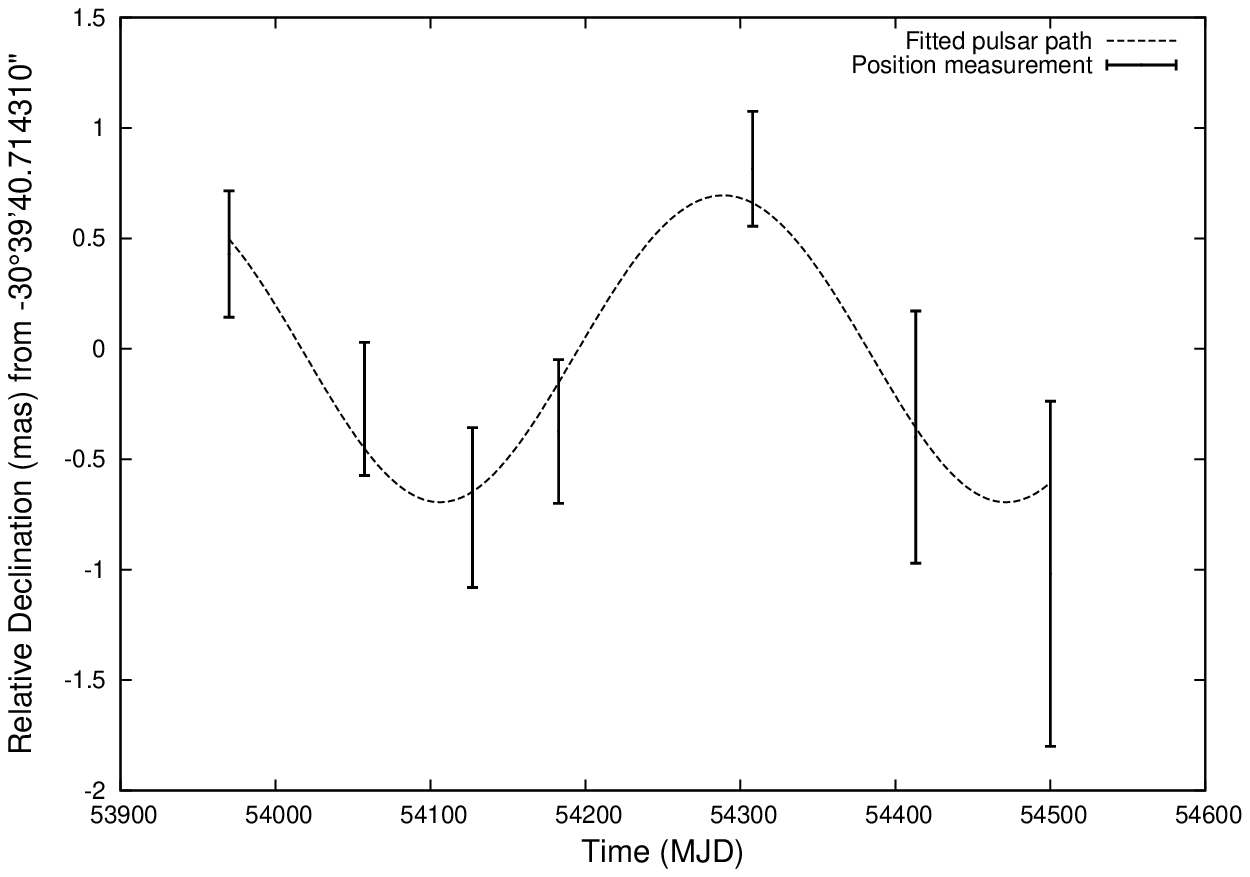} \\
\end{tabular}
\caption{Motion of \pfour, with measured positions overlaid on the best fit. From top to bottom; 
motion in declination vs right ascension, motion in right ascension vs time with 
proper motion subtracted, and motion in declination vs time with proper motion subtracted.}
\label{fig:0737fit}
\end{center}
\end{figure}

\begin{figure}
\begin{center}
\begin{tabular}{c}
\includegraphics[width=0.45\textwidth]{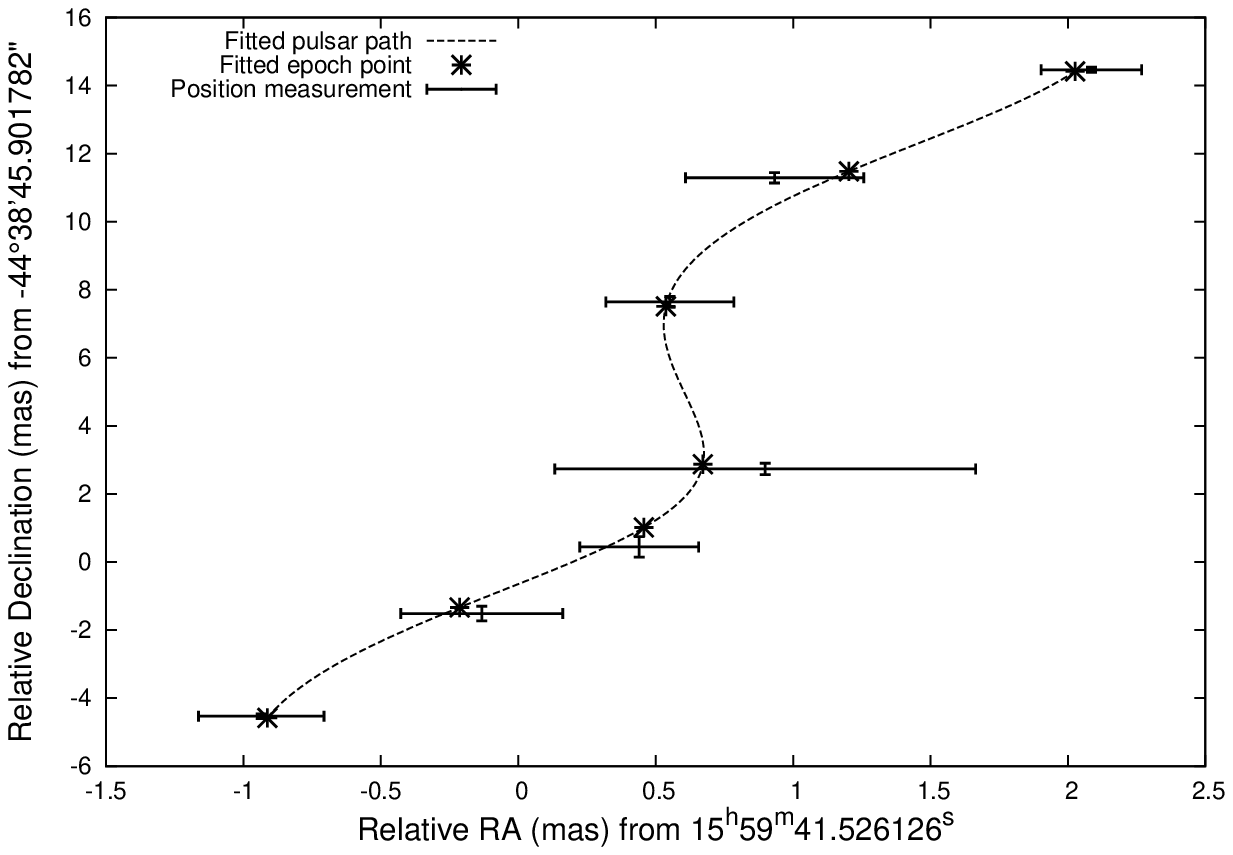} \\
\includegraphics[width=0.45\textwidth]{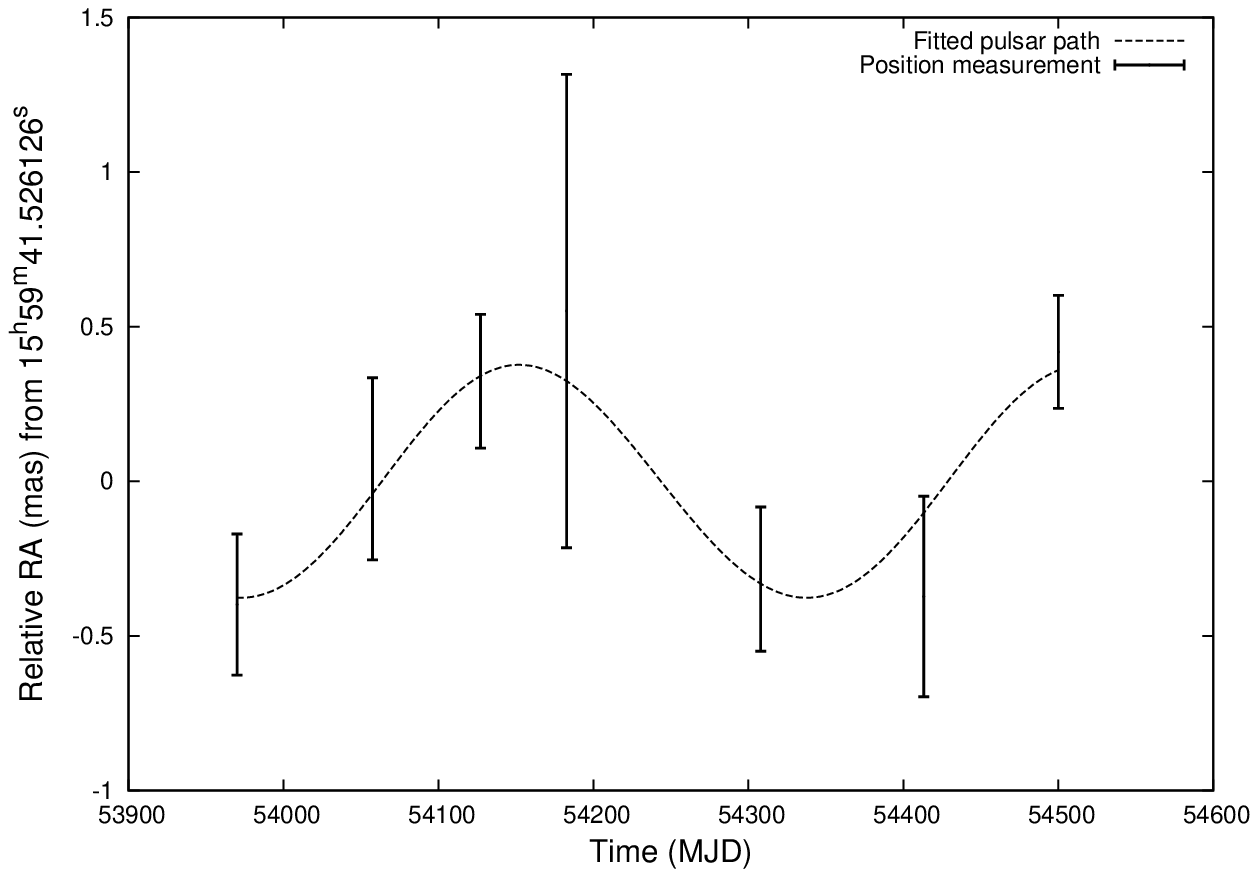} \\
\includegraphics[width=0.45\textwidth]{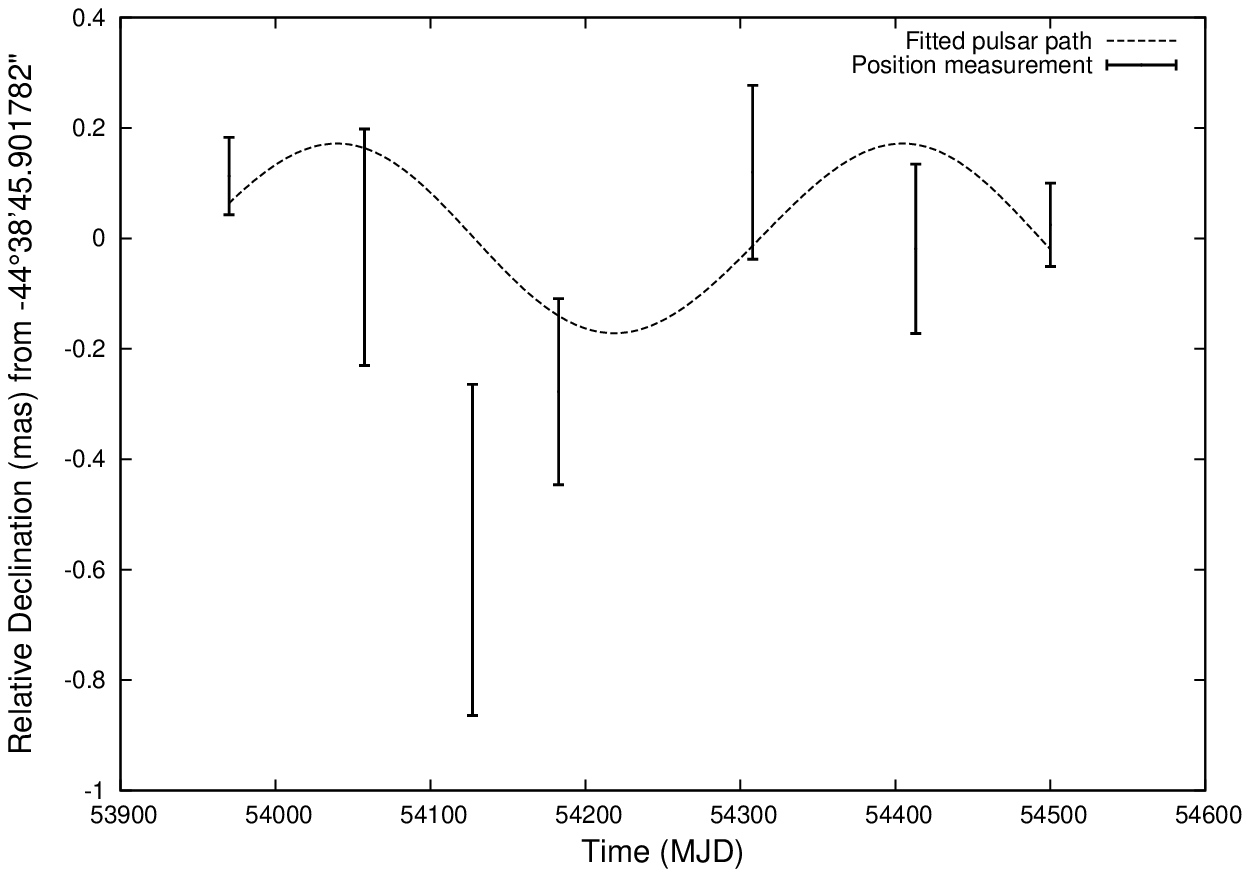} \\
\end{tabular}
\caption{Motion of \pfive, with measured positions overlaid on the best fit. From top to bottom; 
motion in declination vs right ascension, motion in right ascension vs time with 
proper motion subtracted, and motion in declination vs time with proper motion subtracted.}
\label{fig:1559fit}
\end{center}
\end{figure}

Figures~\ref{fig:0108fit}--\ref{fig:2145fit} plot the results for each pulsar, showing
the astrometric fit overlaid on the measured positions with 1$\sigma$\ error bars.
In each figure, the pulsar position is shown in right ascension and declination in
the top panel, while the middle and lower panel show right ascension plotted against 
time and declination plotted against time respectively.  In the lower two panels, the best--fit
proper motion has been subtracted for clarity.  

As discussed in \citet{deller08a},
the optimal weighting for visibility data depends on the strength of the pulsar and the
magnitude of systematic errors.  Following the guidelines developed in \citet{deller08a}, we 
used equally--weighted visibility data when the target could be detected in a single epoch using this weighting
with a signal to noise ratio (SNR) greater than $\sim$10; we have used sensitivity--weighted visibilities
when this criterion could not be met.   For the strongest pulsars, we have
used uniform weighting when imaging to maximise resolution, but natural weighting was
used for weaker pulsars where the use of uniform weighting has an adverse effect
on single--epoch SNR.  These choices are listed in Table~\ref{tab:allresults}.

Along with the recent results of \citet{chatterjee09a}, our detection of seven southern hemisphere 
pulsar parallaxes increases the total number of
VLBI pulsar parallaxes in the south to 12, and the total number of published VLBI 
pulsar parallaxes to 35.  Based on the results presented here, it is possible to estimate the 
number of southern pulsars which remain accessible to the LBA for future parallax studies.
The parallax error of 140 $\mu$as obtained for \pfour, which possesses an ungated 1600
MHz flux of only 2-3 mJy, shows that LBA parallax measurements for moderately faint pulsars
are possible to distances of several kpc.  A search of the ATNF pulsar 
catalogue\footnote{http://www.atnf.csiro.au/research/pulsar/psrcat/} reveals that there are 
over 30 suitable pulsars (declination $<0$, 1400 MHz flux $> 2.5$ mJy and DM distance
$\leq 2$ kpc) that have not previously been the subject of VLBI astrometry.  Future improvements 
to VLBI instrumentation should rapidly make even more pulsars accessible.
Thus, future parallax observations with the LBA could and should continue to redress
the imbalance of measured pulsar parallaxes at southern declinations.

The implications of the measured parallax and proper motion
for each pulsar are discussed below in \S\ref{sec:discussion}, along with the implications
of the inferred error distribution for distance estimates based upon Galactic electron models
and pulsar DMs.

\section{Discussion}
\label{sec:discussion}
\subsection{PSR J0108--1431}
\label{sec:discussion:0108}
\pone\ was first reported by  \citet{tauris94a}, and was postulated to be the nearest observed
radio pulsar -- its DM of 2.38 remains the lowest of any known pulsar.  In the TC93 
Galactic electron model, \pone\ was predicted to lie 130 pc from Earth -- in the newer
NE2001 model \citep{cordes02a}, the distance was revised to 184 pc, which if correct would 
double the pulsar luminosity, to a value of 300 $\mu$Jy kpc$^{2}$\ at 400 MHz (the lowest known 
value is that of PSR J0006+1834, which at a DM--derived 
distance of 700 pc has a predicted 400 MHz luminosity of 100 $\mu$Jy kpc$^{2}$).

Figure~\ref{fig:0108fit} shows the fitted and measured positions for \pone.  As shown in 
Table~\ref{tab:allresults}, the measured distance of  $240_{-61}^{+124}$\,pc is just 
consistent with the NE2001 distance, but inconsistent at 95\% confidence with the earlier
TC93 distance.  It confirms that \pone\ is more distant than originally suspected, and while
its apparent luminosity is still low, it is no longer remarkably so.  In fact, 5 other pulsars 
(PSR J0006+1834, PSR J1829+2456, PSR J1918+1541, PSR J2015+2524, and PSR J2307+2225) 
have lower apparent 400 MHz luminosities than the revised value of 510 $\mu$Jy kpc$^{2}$\ for \pone.  

These observations represent the first measurements of the proper motion of \pone, as well
as the first distance measurement.
Knowing the distance and transverse velocity of \pone\ makes it possible to calculate the
Shklovskii correction to its period derivative $\dot{P}$, which is given by:

\begin{equation}
\frac{\dot{P}_{\mathrm{Shk}}}{P} = \frac{\mu^{2}D}{c}
\end{equation}

\noindent where $D$ is the pulsar distance, $\mu$ is proper motion and $c$ is the speed of light.
For \pone, the Shklovskii correction to the observed $\dot{P}$ is $1.37\times10^{-17}$, or 
18\% of the observed value of $7.704\times10^{-17}$\ \citep{hobbs04a}.
This correction to $\dot{P}$ allows a more accurate estimate of the pulsar's 
spin--down luminosity $\dot{E} = 4\pi^{2}I\dot{P}/{P}^{3}$ and characteristic age
$\tau_{c} = P/((n-1)\dot{P})$, where $I$ is the neutron star 
moment of inertia (estimated to be $10^{45}$ g cm$^{2}$) and 
$n$ is the braking index, equal to 3 for magnetic dipole braking in a vacuum.
For \pone, the value of $\dot{E}$ is revised from $5.8\times10^{30}$ erg s$^{-1}$ to 
to $4.8\times10^{30}$ erg s$^{-1}$, and the value of $\tau_{c}$ is revised from
$166\times10^{6}$ years to $200\times10^{6}$ years.
  
\pone\ has recently been detected in x--rays \citep{pavlov09a}, prompting a re--examination of
archival optical data and a subsequent optical detection \citep{mignani08a}.  The VLBI position
and proper motion presented here confirms that the optical detection of \citet{mignani08a} is
coincident with the pulsar position at the time of the observation. 
 The x--ray emission is consistent with a non--thermal spectrum of presumed
magnetospheric origin, while the optical detection was consistent with
thermal emission from the bulk of the neutron star surface.  For both the x--ray and optical detections,
the revised distance measure implies a greater energy output in these wavebands, although the 
low precision of the parallax measurement means the luminosities remain relatively poorly
constrained.

For the x--ray waveband, the combination of revised distance measure and spin--down luminosity
implies that \pone\ actually converts 1.5$^{+2.0}_{-0.7}$\% of its spin--down luminosity into x--rays, rather than the
value of 0.4\% reported in \citet{pavlov09a}.  This represents the highest known x--ray 
conversion efficiency
of any pulsar -- the distance and hence x--ray conversion efficiency of the previously most efficient 
pulsar (\pthree) has been revised dramatically downwards by the measurements presented here. 
\pthree\ is discussed further in \S\ref{sec:discussion:0630}.

In the optical waveband, the upwardly revised luminosity requires either the addition of a large non--thermal
component to the emission, or if the emission is solely thermal, a significant 
increase in the neutron star surface temperature.  Any non--thermal optical emission would be generated
at a much higher efficiency than other comparable old neutron stars, although the ratio of optical efficiency to 
x--ray efficiency would be consistent with that found by \citet{zharikov04a}.  Unlike PSR B0950+08, the next 
oldest pulsar with multiwavelength data, the slope of a putative power--law component could not
be consistent with that inferred from x--ray observations \citep{mignani08a}.  However,
single power law fits are not ubiquitous amongst younger pulsars, and the contribution of non--thermal optical 
emission to the spectrum of \pone\ cannot be discounted.

If negligible magnetospheric optical emission is assumed, the implied surface temperature for \pone\ far exceeds
that predicted for a neutron star of age $200\times10^{6}$\,yr (Shklovskii--corrected).  Scaling the surface temperature
obtained by \citet{mignani08a} by the increased optical luminosity at the best--fit distance of 240 pc yields 
a surface temperature of $(2.9\pm0.5)\times10^{5}$\,K;  taking the 95\% confidence lower limit for distance (143 pc) 
and the lower temperature limit given in \citet{mignani08a} yields a minimum surface temperature of $8.5\times10^{4}$\,K.
According to standard neutron star cooling models, the maximum temperature of an isolated neutron
star of age $200\times10^{6}$ yr should be $3\times10^{4}$\,K if no reheating occurs \citep[see Figure 3 of][]{schaab99a}.
To obtain a temperature above the the 95\% lower limit, the neutron star could be no older than 
$10\times10^{6}\,$yr -- an unlikely prospect, as it would require the pulsar's birth spin period to be at least
770 ms, much higher than typically assumed values \citep[see e.g.][]{faucher-giguere06a}.
Improved optical observations are needed to confirm or deny the presence of magnetospheric optical emission,
but these results are supportive of the presence of significant reheating for old, isolated neutron stars.  

\subsection{PSR J0437--4715}
\label{sec:discussion:0437}
The results obtained for PSR J0437--4715 have already been discussed in \citet{deller08b}, and
a brief summary is presented here.  The fit to the observed pulsar positions is shown in 
Figure~\ref{fig:0437fit}.  The pulsar distance of $156.3\pm1.3$\ pc which is derived from our
VLBI astrometry is the most accurate pulsar distance (in both absolute and fractional distance)
obtained for any pulsar to date.  The comparison of this distance with the ``kinematic distance"
derived by \citet{verbiest08a} from the measurement of the rate of change of binary period yields 
a limit on the maximum apparent acceleration of PSR J0437--4715 along the line of sight to the 
solar system.  As shown in \citet{deller08b}, this can in turn be used to place limits on the maximum
rate of change of Newton's gravitational constant $G$ with time 
($\dot{G}/G = (-5 \pm 26)\times 10^{-13}$\ yr$^{-1}$, 95\% confidence), 
as well as the maximum mass of any undetected massive planets orbiting the solar system
(no Jupiter--mass planets within 226 AU of the solar system barycentre, 95\% confidence over
50\% of the sky), and the maximum amplitude of the characteristic strain spectrum of the 
stochastic gravitational wave background (an amplitude of $1.1\times10^{-13}$\ at a period of one 
year produces inconsistent distances in 95\% of simulations).

The covariance between proper motion and parallax for \ptwo\ was the second highest of the pulsars in
our sample after \psix.  Since considerable time has passed since the end of our observing program, the
addition of several new epochs could considerably improve the proper motion and parallax accuracy
for \ptwo.  Through simulations, we have estimated that three additional epochs over a year would reduce 
the parallax uncertainty for \ptwo\ by a further factor of two. This extension to our observing program is 
the subject of a currently active LBA proposal.

\subsection{PSR J0630--2834}
\label{sec:discussion:0630}
\pthree\ is a middle--aged pulsar which has been observed extensively as part of scintillation
studies \citep[see e.g.][]{cordes86b, bhat99a}, 
and would be undistinguished if not for its remarkable apparent luminosity
in x--rays.  \citet{becker05a} observed \pthree\ with XMM--Newton and found an x--ray luminosity
of $8.4\times10^{30}$\,erg s$^{-1}$, based on the NE2001 distance of 1.45 kpc. 
This implied that the pulsar was converting
$\sim\,$16\% of its spin--down luminosity into x--rays -- an order of magnitude more than any other
old or middle--aged pulsar.  They suggest that the most likely explanation
is that the distance to the pulsar is over--estimated, a theory which has been proven correct by 
these VLBI observations.  

The measured and fitted positions of \pthree\ are shown in Figure~\ref{fig:0630fit}.  
The measured distance of $332^{+52}_{-40}$\ pc, as shown in Table~\ref{tab:allresults},
means that the actual x--ray conversion efficiency is a much less surprising $0.8^{+0.3}_{-0.2}$\,\%.  It is also
within the $1\sigma$\ error bars of the efficiency derived for \pone\ in \S\ref{sec:discussion:0108}
above.  The best--fit distance of 332 pc is less than a quarter of the DM--derived distance in the 
NE2001 model, and a factor of seven smaller than the distance predicted by the TC93 model.  
This large discrepancy is discussed further in \S\ref{sec:discussion:dm}.

The proper motion of \pthree\ has previously been measured using the VLA by \citet{brisken03b},
and the measured values from this work 
($\mu_{\alpha} = -46.3\pm 1.0$\,mas yr$^{-1}$, $\mu_{\delta} = 21.3 \pm 0.5$\,mas yr$^{-1}$) 
are consistent with, but marginally
more precise, than the VLA results ($\mu_{\alpha} = -44.6\pm 0.9$\,mas yr$^{-1}$, 
$\mu_{\delta} = 19.5 \pm 2.2$\,mas yr$^{-1}$).  The resultant Shklovskii correction is less than 0.04\%
of the observed period derivative ($7.12\times10^{-15}$). 

\subsection{PSR J0737--3039}
\label{sec:discussion:0737}
The results obtained for PSR J0737--3039 have already been discussed in \citet{deller09a}, 
and a brief summary is presented here.
The fit to the observed pulsar positions is shown in Figure~\ref{fig:0737fit}.
The measured distance of $1150^{+220}_{-160}$\,pc is inconsistent with previous DM--based 
estimates of 570 pc (TC93) and 480 pc (NE2001), and is much more significant
than the marginal timing parallax detection of $333^{+667}_{-133}$\,pc \citep{kramer06a}. 

The kinematic information obtained from these measurements can be used to accurately calculate
the Shklovskii term and other kinematic contributions to the binary period derivative of \pfour\ for
the first time, allowing an estimation of the precision to which GR can be tested.  
The measurements reported here reduce the uncertainty on kinematic terms in 
the observed binary period derivative to 0.01\% of the GR contribution, implying that with 
increasing accuracy of timing observations, tests of GR can be made in the strong--field regime to
the 0.01\% level.  The kinematic information also implies that the transverse velocity of
\pfour\ in the local standard of rest (LSR) is only $9^{+6}_{-3}$\,\kms.  Given that the system
must have survived two supernovae undisrupted, this low velocity is surprising, and lends 
support to models of formation of the system in which the neutron stars receive low or no
kicks at birth \citep[e.g.][]{piran05a}.

Finally, the revised distance measure helps to clarify
the origin of the bulk of the x--ray emission in 
the \pfour\ system.  A magnetospheric origin for the x--radiation (from pulsar A) had been considered
a strong possibility, but the hydrogen column density calculated from fits to the x--ray data had appeared 
to be unreasonably high, given the presumed location of \pfour\ in the Gum nebula HII region \citep{possenti08a}.  
The revised distance measure places the pulsar beyond the Gum nebula and makes the 
calculated hydrogen column density consistent with expectations.

\subsection{PSR J1559--4438}
\label{sec:discussion:1559}
The results obtained for PSR J1559--4438 have already been discussed in \citet{deller08a},
and a brief summary is presented here. 
The fit to the observed pulsar positions is shown in Figure~\ref{fig:1559fit}.
At a distance of $2600^{+690}_{-450}$\,pc, \pfive\ is one of the most distant pulsars with
a parallax measured using VLBI.  
The measured distance is consistent with the NE2001 prediction 
\citep[2350 pc;][]{cordes02a}, which differed considerably from the earlier 
TC93 distance estimate of 1580 pc.  
It is also consistent with the lower distance estimate of $2.0 \pm 0.5$ kpc made using HI
line absorptions by \citet{koribalski95a}. 
The measured values of
proper motion ($\mu_{\alpha} = 1.52 \pm 0.14$\,mas yr$^{-1}$, 
$\mu_{\delta} = 13.15 \pm 0.05$\,mas yr$^{-1}$) 
are consistent with the VLA observations of \citet{fomalont97a}, who measured
$\mu_{\alpha} = 1 \pm 6$\,mas yr$^{-1}$, $\mu_{\delta} = 14 \pm 11$\,mas yr$^{-1}$.

With an accurate proper motion now calculated, the position angle of the proper motion
for \pfive\ can be compared to the position angle of the emission polarisation, which tests
the alignment of the rotation and velocity vector, as described by \citet{johnston07a}.
In the case of \pfive, the position angle of the proper motion ($6.6 \pm 0.6$\degrees) is not
significantly aligned parallel with or perpendicular to the polarisation position angle ($71\pm3$\degrees).  
However, as noted in \citet{deller08a}, the polarisation profile of \pfive\ (as shown in
Figure~5 of \citealt{johnston07a}) is complicated, and it is possible that the original determination of
the magnetic field orientation from the polarisation position angle was incorrect.  

\subsection{PSR J2048--1616}
\label{sec:discussion:2048}
\begin{figure}
\begin{center}
\begin{tabular}{c}
\includegraphics[width=0.45\textwidth]{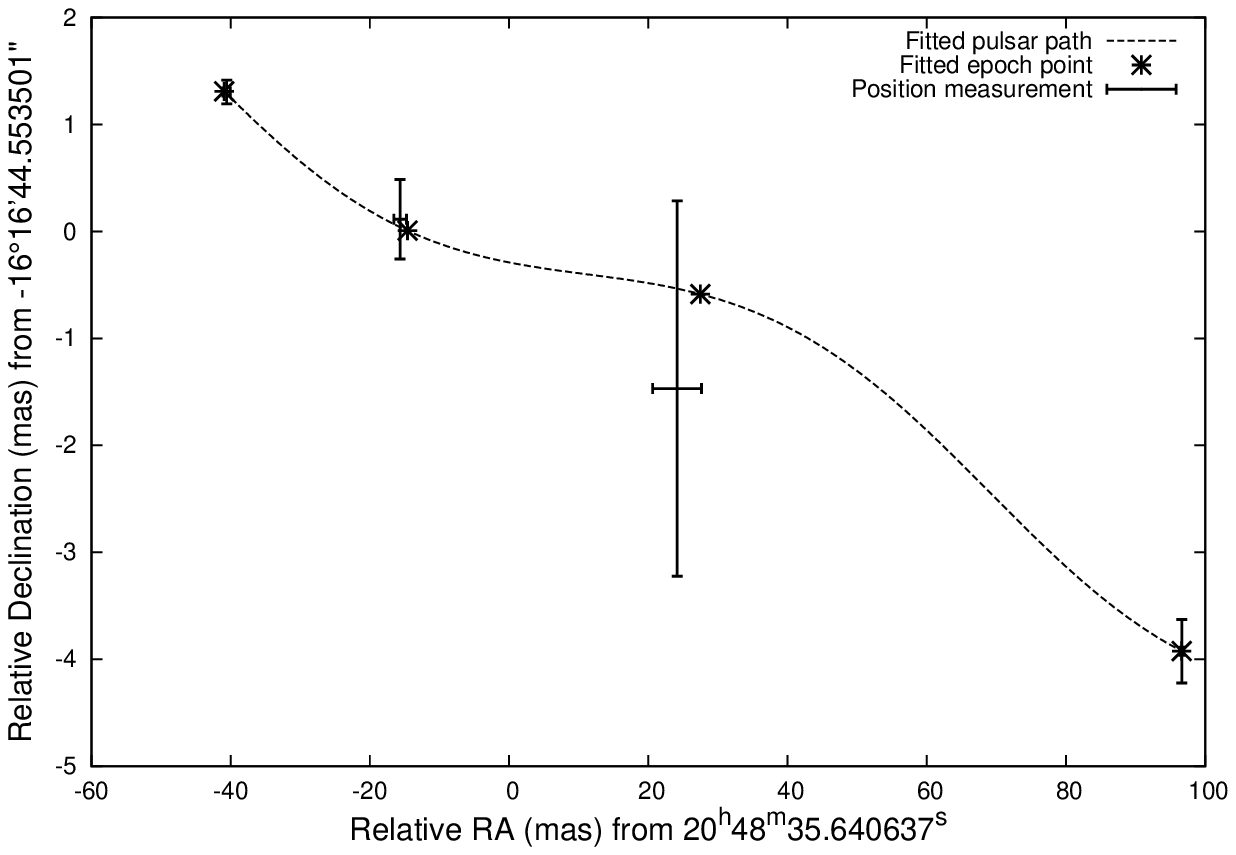} \\
\includegraphics[width=0.45\textwidth]{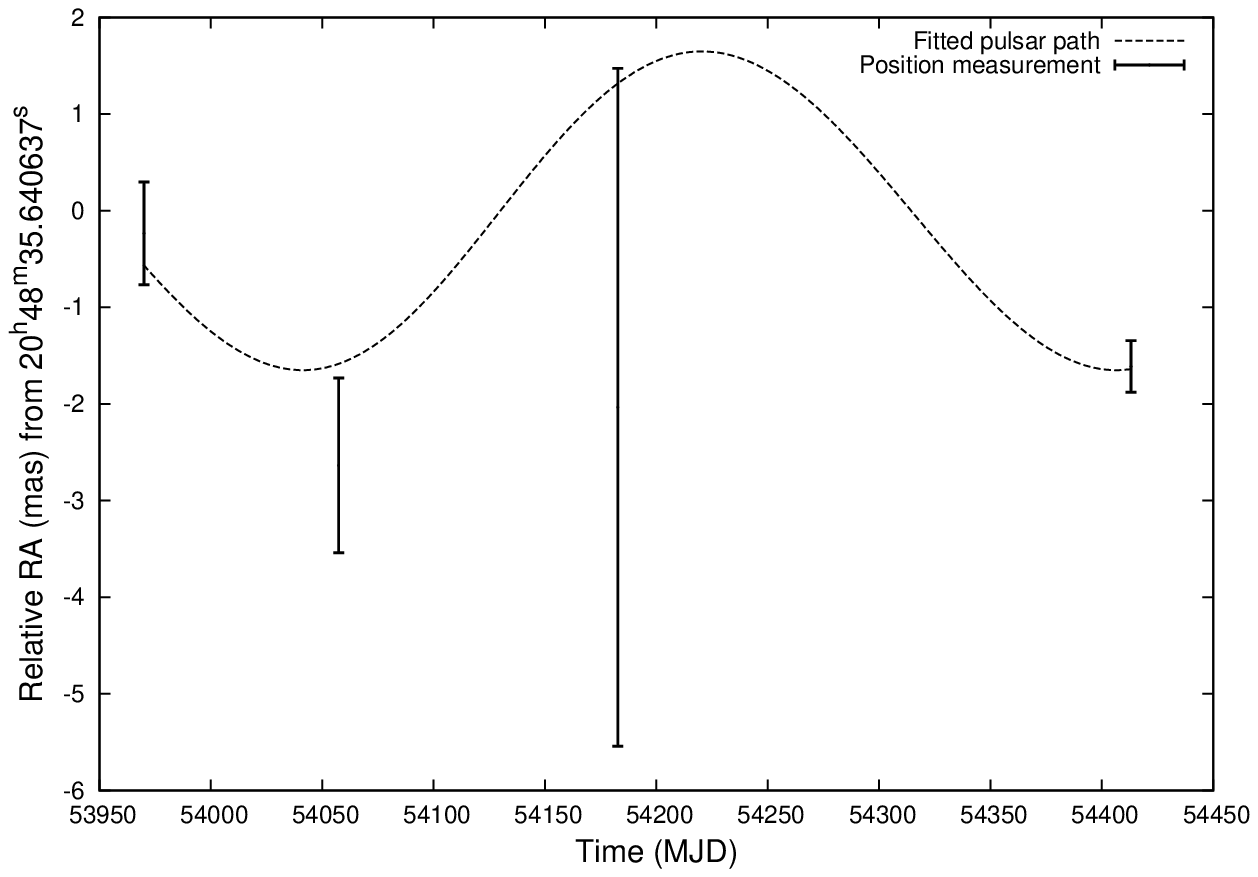} \\
\includegraphics[width=0.45\textwidth]{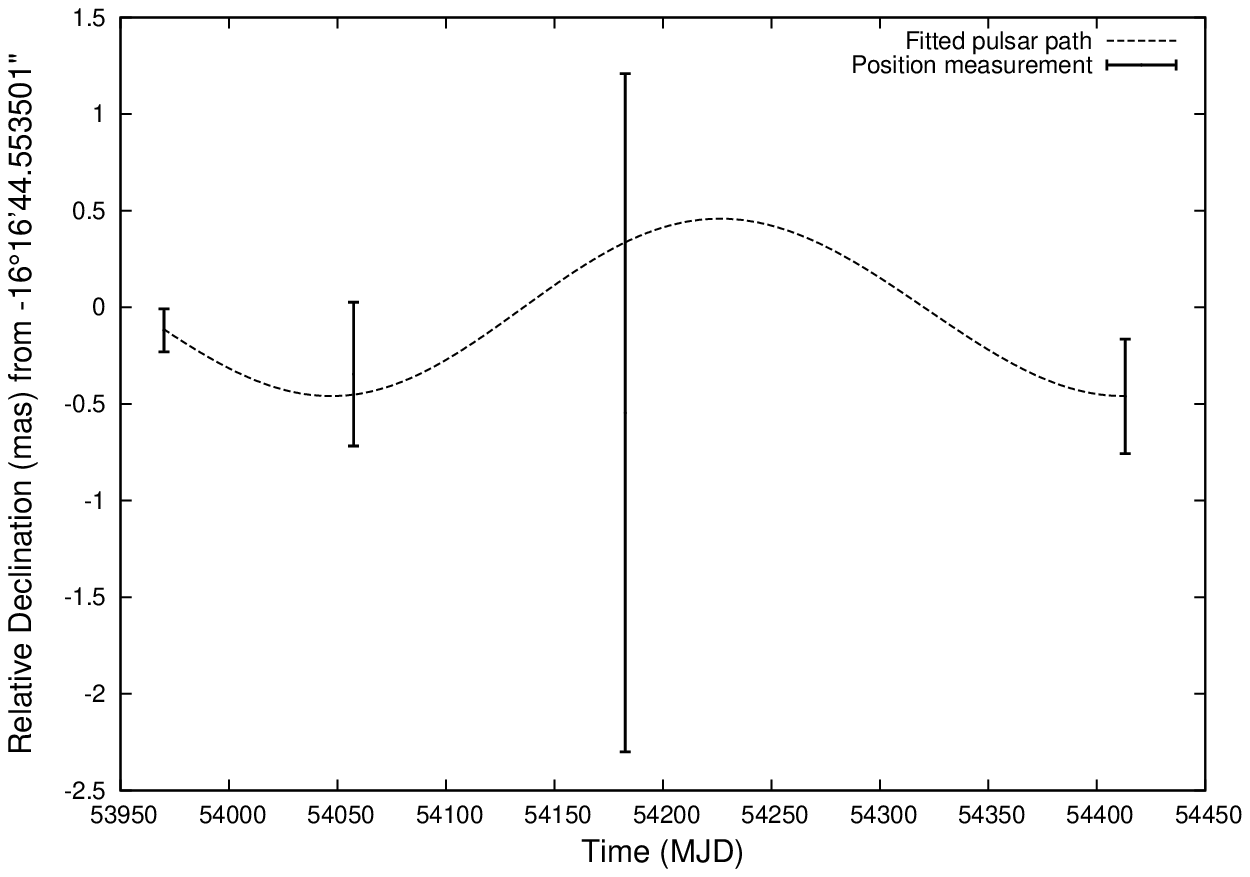} \\
\end{tabular}
\caption{Motion of \psix, with measured positions overlaid on the best fit. From top to bottom; 
motion in declination vs right ascension, motion in right ascension vs time with 
proper motion subtracted, and motion in declination vs time with proper motion subtracted.}
\label{fig:2048fit}
\end{center}
\end{figure}

\begin{figure}
\begin{center}
\begin{tabular}{c}
\includegraphics[width=0.45\textwidth]{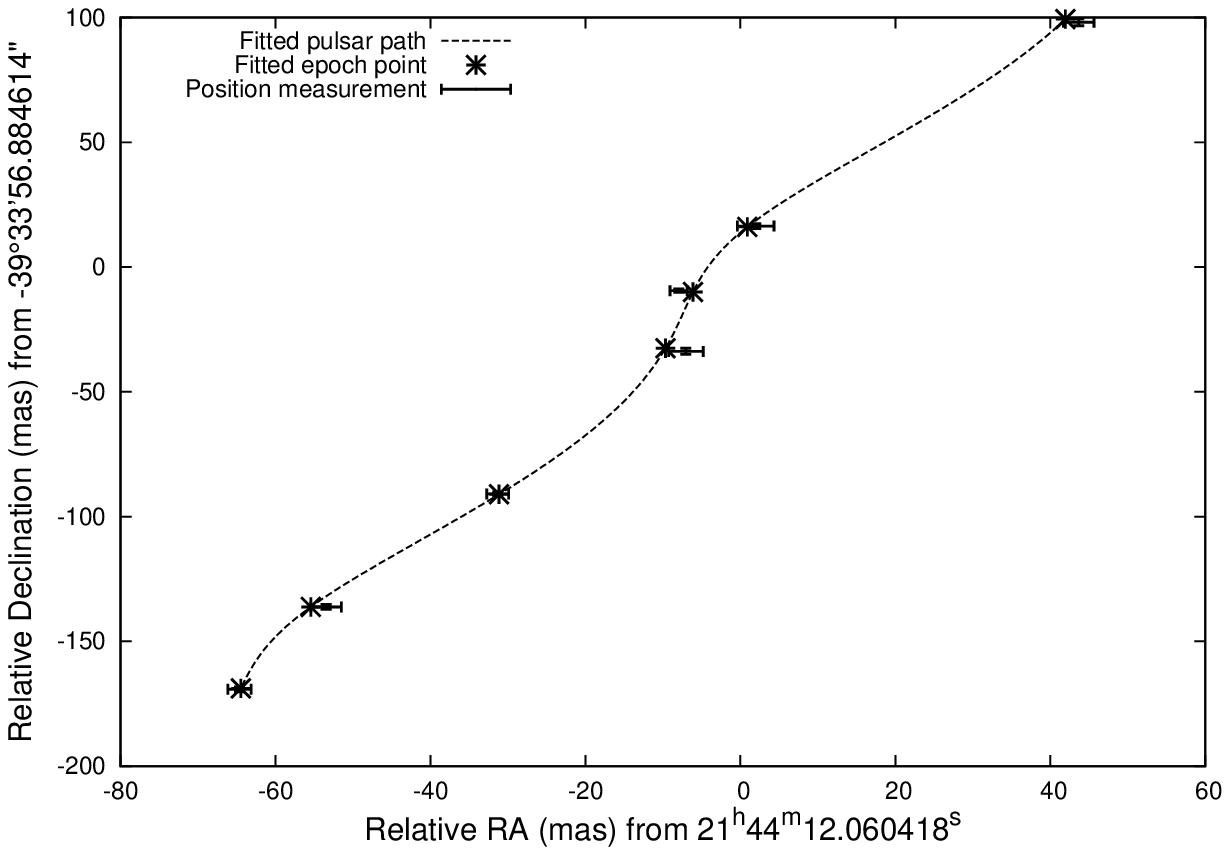} \\
\includegraphics[width=0.45\textwidth]{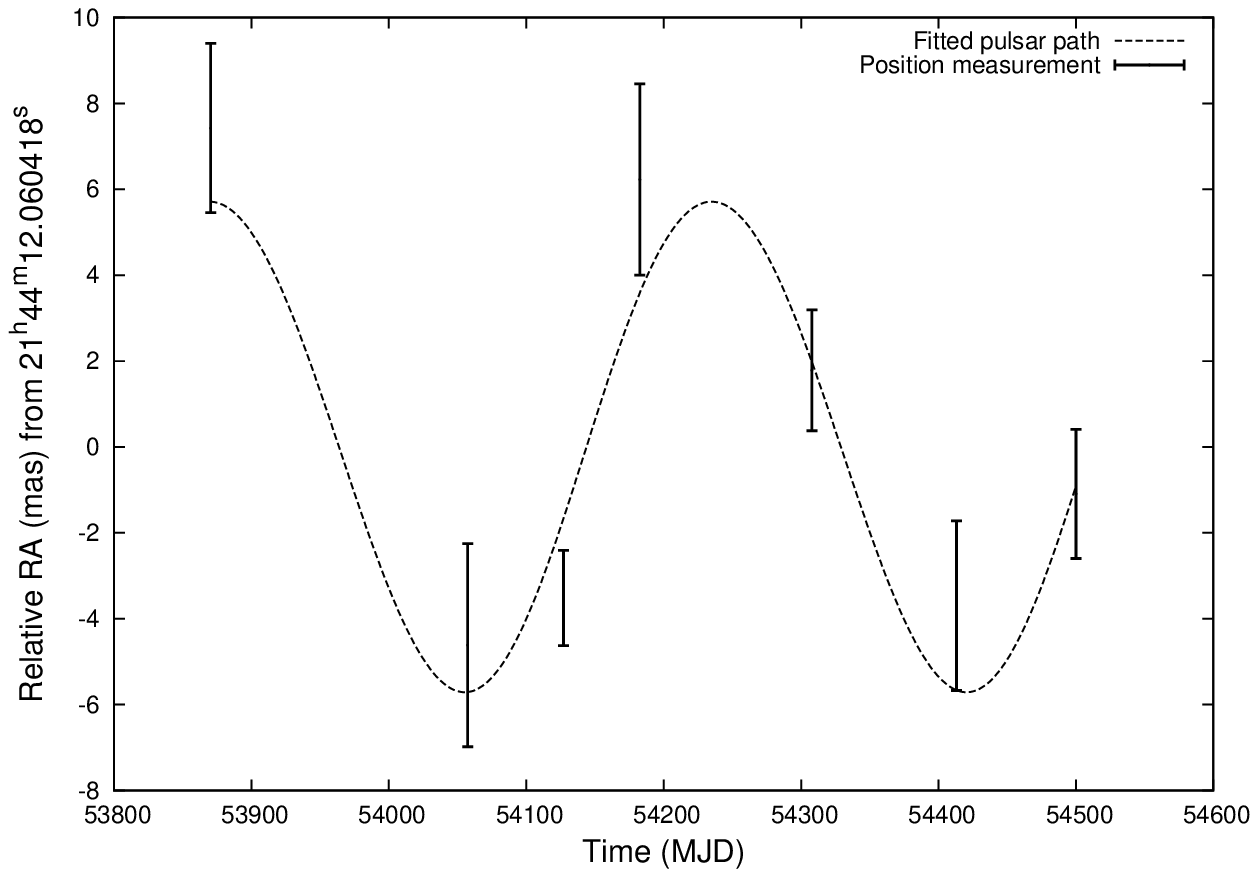} \\
\includegraphics[width=0.45\textwidth]{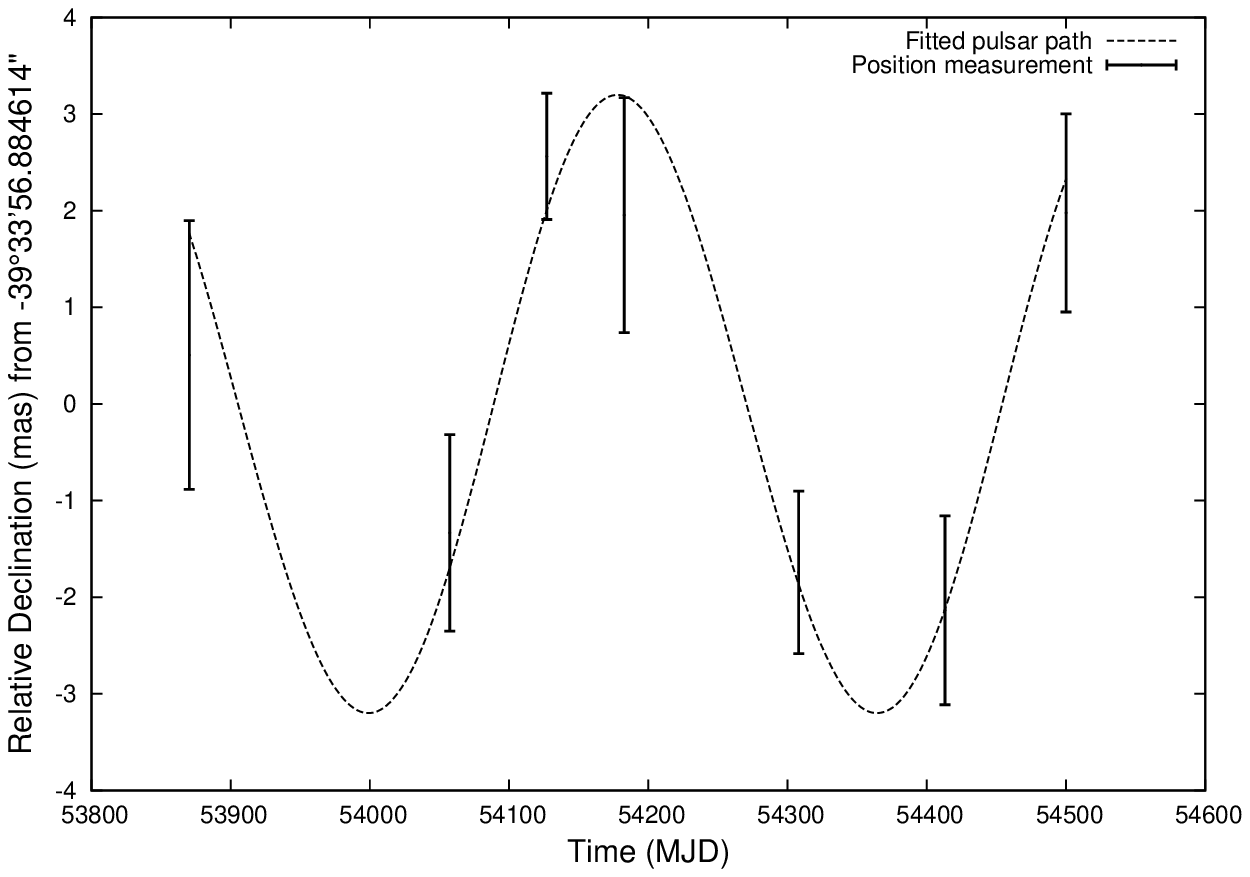} \\
\end{tabular}
\caption{Motion of \pseven, with measured positions overlaid on the best fit. From top to bottom; 
motion in declination vs right ascension, motion in right ascension vs time with 
proper motion subtracted, and motion in declination vs time with proper motion subtracted.}
\label{fig:2144fit}
\end{center}
\end{figure}

\begin{figure}
\begin{center}
\includegraphics[width=0.45\textwidth]{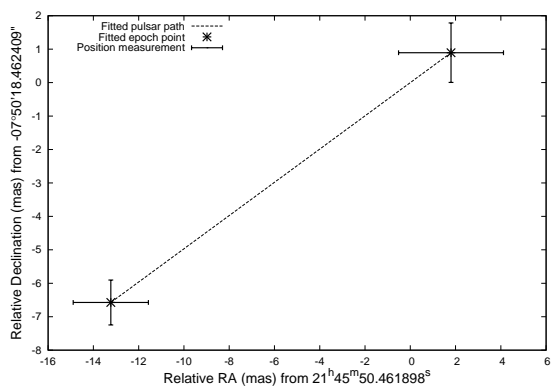}
\caption{Motion of \peight\ in right ascension plotted against motion in declination, 
with measured positions overlaid on the best fit. 
Since only two significant measurements were made, the parallax of the pulsar was held 
fixed at zero, and the fit to position and proper motion has no free parameters.  As the parallax
was fixed at zero, the plots of motion in right ascension vs time and declination vs time which were
shown for other pulsars are not shown here.  Variation
of the parallax between zero and two mas made an insignificant 
($\lesssim100\,\mu$as yr$^{-1}$) difference to fitted proper motion.}
\label{fig:2145fit}
\end{center}
\end{figure}

\psix\ was the second ``technique check" source included in the observing program, after \pfive.  It
is less bright than \pfive, but was predicted to be somewhat closer (640 pc in the TC93 model,
560 pc in the NE2001 model).  As shown in Table~\ref{tab:allresults}, whilst a parallax
was measured for \psix, it was not significant (1.9$\sigma$).  Nevertheless, the best--fit distance of
580 pc is consistent with the DM--based distance estimates, and an accurate measurement
of proper motion was made.  Figure~\ref{fig:2048fit} shows the fitted and measured positions
of \psix.

It is apparent from Figure~\ref{fig:2048fit} that the large position error of the third epoch (MJD 54182) is
the primary reason for the poorly constrained fit.  Equipment failure at the Mopra telescope and limited
time on the Tidbinbilla telescope during this epoch reduced the number of telescopes on--source during
observations of \psix\ to three, and thus the large errors are unsurprising.  The covariance between 
proper motion and parallax was the largest for \psix, and hence one or more additional 
position measurements, preferably close to the appropriate parallax extrema, could have significantly
improved the parallax determination.

A considerably more accurate parallax for \psix\ has recently been 
reported by \citet{chatterjee09a}, who observed \psix\ along with 13 other pulsars using the 
VLBA.  The parallax of $1.05^{+0.03}_{-0.02}$\ mas obtained by \citet{chatterjee09a} is consistent
with our results (1.71$\pm$0.91 mas), but the proper motion values differ significantly 
($\mu_{\alpha} = 113.16 \pm 0.02$ mas yr$^{-1}$, $\mu_{\delta} = -4.60^{+0.28}_{-0.23}$ mas yr$^{-1}$ 
for \citealt{chatterjee09a}; 
$\mu_{\alpha} = 114.24 \pm 0.52$ mas yr$^{-1}$, $\mu_{\delta} = -4.03 \pm 0.24$ mas yr$^{-1}$ 
from our observations).
Combining the epoch position fits from both observing campaigns and allowing a constant offset between the LBA and
VLBA epochs yields a fit which is consistent with the VLBA--only values, with similar errors.  This is 
unsurprising since the VLBA epochs, with much lower errors, dominate the fit.  Refitting the LBA--only data after
fixing the parallax to the VLBA--derived value of $1.05^{+0.03}_{-0.02}$\ mas removes the impact of the covariance with 
parallax and yields $\mu_{\alpha} = 113.94 \pm 0.32$ mas yr$^{-1}$ and $\mu_{\delta} = -4.30 \pm 0.10$ mas yr$^{-1}$; 
closer to the VLBA--only values, but still differing at the 2$\sigma$ level in right ascension.

Whilst both observing campaigns utilized the same reference source (J2047--1639), they sampled different time periods
(2002 -- 2005 for \citealt{chatterjee09a}, 2006 -- 2008 for these observations) and the VLBA observations
utilized an in--beam calibrator.  The primary calibrator is slightly resolved, as shown in 
Figure~\ref{fig:cals2}, and thus imperfect modeling of its (possibly time varying) structure 
is a plausible explanation for the proper motion discrepancy.  The problem of modeling the structure of resolved 
calibrator sources is particularly acute for the LBA, due to the limited $uv$ coverage.

\subsection{PSR J2144--3933}
\label{sec:discussion:2144}
\pseven\ was discovered in the Parkes Southern Pulsar Survey \citep{manchester96a}, and 
was initially misidentified with a period of 2.83 seconds.  \citet{young99a} reported that
the true period is in fact 8.5 seconds, making \pseven\ the longest period radio pulsar known.
\pseven\ lies below the traditionally assumed pulsar ``death line" \citep[see e.g.][]{chen93a}, 
and hence its true luminosity is extremely important for models of pulsar emission and evolution.  
The discovery of \pseven\ prompted alternative models of pulsar emission in which 
long period pulsars such as \pseven\ could remain luminous in the radio \citep[e.g.][]{zhang00a}.
\pseven\ possesses a steeper spectral index than average ($-2.4$), 
and while several other pulsars are less
luminous at 400 MHz, \pseven\ is the least luminous pulsar known at 1400 MHz 
(24 $\mu$Jy kpc$^{2}$\ at assumed distance of 180 pc; average 1400 MHz flux density of 0.75 mJy 
calculated from archival Parkes observations).

The fitted and measured positions of \pseven\ are shown in Figure~\ref{fig:2144fit}.  As shown in
Table~\ref{tab:allresults}, a highly significant parallax was detected, corresponding to a 
distance of $165^{+17}_{-14}$\,pc.  Given the generally assumed errors on DM distances,
this is consistent with the TC93 value (180 pc), but not the NE2001 value (264 pc).  This
confirms that the apparent radio luminosity of \pseven\ is extremely low -- 15\% lower than the
previously assumed value of 24 $\mu$Jy kpc$^{2}$.

The proper motion measurement allows for the first time a calculation of the Shklovskii correction
for the period derivative of \pseven\ -- the Shklovskii contribution to $\dot{P}$ is $9.4\times10^{-17}$,
or approximately 19\% of the observed $\dot{P}$\ of $4.96\times10^{-16}$.  The true spin--down 
luminosity of \pseven\ is thus further reduced from the assumed value of 
$3.2\times10^{28}\,$erg s$^{-1}$\ to $2.6\times10^{28}\,$erg s$^{-1}$, the smallest known 
spin--down luminosity of any pulsar.  PSR J0343--3000, discovered in
the Parkes High--Latitude survey by \citet{burgay06a}, has the next lowest spin--down luminosity,
which is a factor of 5 higher than the revised value for \pseven.  In addition, this revision to the 
$\dot{P}$\ value for \pseven\ places it even further past the assumed pulsar ``death line".

As noted by \citet{young99a}, the low luminosity and narrow
beaming fraction ($\sim0.01$) of \pseven\ imply that many such objects exist in the Galaxy.  
These measurements confirm
that the true distributions of pulsar spin--down luminosities and radio luminosities do extend down to
very low levels, which is important for the generation of synthetic pulsar catalogues and predictions
of discovery rates for instruments such as the Square Kilometre Array 
(SKA\footnote{http://www.skatelescope.org}) and LOFAR\footnote{http://www.lofar.org/}.  
Assuming the steep spectral index of \pseven\ is typical for similar long-period, 
low--luminosity objects, and that the Galactic population of such objects can be estimated
from \pseven\ alone, more than 100,000 similar pulsars may inhabit the Galaxy \citep{young99a}.

\subsection{PSR J2145--0750}
\label{sec:discussion:2145}
\peight\ is a binary pulsar with spin period 16.05 ms and orbital period 6.84 days, which was 
discovered by \citet{bailes94a}.   Optical observations by \citet{bell95a} tentatively identified the 
expected white dwarf companion to \peight, which was confirmed by \citet{lundgren96a}.  
Timing astrometry using a 10 year dataset by \citet{lohmer04a} measured a 
parallax of $2.0 \pm 0.6$\,mas,
and a proper motion of $14.1\pm0.4\,$mas yr$^{-1}$, corresponding to a transverse velocity of 
$33\pm9$\,\kms.  However, \citet{hotan06a} reported no detection of parallax and an upper
limit of 0.9 mas, at 95\% confidence, albeit with a shorter (2.5 year) dataset, in addition to 
a proper motion measurement of $13.5\pm6.0\,$mas yr$^{-1}$.  
A previous attempt has been made to measure the VLBI parallax of this system 
by \citet{brisken02a} using the VLBA, but the pulsar was not detected and dropped
from the observing program.  Accordingly, the distance to this system is still controversial and 
confirmation of a previous result is keenly sought.

From the four observations made of \peight\ during this observing program, significant detections
were made on only two occasions.  As was the case for the VLBA program of \citet{brisken02a}, 
it is believed that refractive scintillation is the major cause of the non--detections, as the pulsar's flux 
varies significantly over long timescales.  Although measurement of the parallax of \peight\ was
thus not possible, by holding parallax fixed at zero mas it was possible to measure a proper motion
of $\mu_{\alpha} = -15.4 \pm 2.1\,$mas yr$^{-1}$, $\mu_{\delta} = -7.7 \pm 0.80\,$mas yr$^{-1}$.  
Varying the parallax between zero and two mas results in a change in 
proper motion of less than 0.1 mas yr$^{-1}$, much smaller than the formal errors. As the proper
motion and position have been derived from only two measurements, there are no degrees
of freedom to the fit and no estimate of systematic errors, 
and so the result should be treated with some caution.  The 
fitted and measured positions of \peight\ are shown in Figure~\ref{fig:2145fit}.

The observed total proper motion of $17.2\pm2.2$\,mas yr$^{-1}$\ is consistent with the value of
$13.5\pm6.0\,$mas yr$^{-1}$ obtained by \citet{hotan06a}, but only consistent at the 2$\sigma$ level
with the value of $14.1\pm0.4\,$mas yr$^{-1}$ obtained by \citet{lohmer04a}\footnote{Since 
\peight\ lies close to the ecliptic, timing observations obtain large covariances between proper 
motion in right ascension and declination, and so the comparison of total proper motion is more useful.}.
As these VLBI results are derived from only two measurements, and there is no way of quantifying 
the systematic error, it is likely that the lower value of proper motion derived from timing 
($\sim$14 mas yr$^{-1}$) is more accurate for this pulsar.

\subsection{Comparison to DM distance models}
\label{sec:discussion:dm}
Galactic electron density distribution models are extremely important for all fields of pulsar research,
since their prediction of pulsar distances based on DM has the potential to bias estimates
of luminosity, velocity, and various timing terms for the vast majority of pulsars without an
independent distance constraint.  Hence, continual improvement of these models, and characterisation
of errors in the existing models (and the impact of these errors) is a crucial task.

The seven pulsar parallaxes presented here make a substantial 
addition to the 44 pulsars with published parallaxes,
and hence a review of the accuracies of the TC93 and NE2001 models is timely.  For the 
comparison presented below, only pulsars with a parallax more significant than $2\sigma$\ are
considered, and where multiple parallax measurements exist, only the most accurate is used.
This leads to a sample size of 41 (33 from VLBI and 8 from timing).
It is appropriate to note 
that large distance errors are likely to be over--represented in existing astrometric results. 
One reason is the selection effect of anomalous pulsars being chosen for study, which
was certainly the case for the \pthree\ system.
Another potential factor is that most astrometric distance determinations to date have been for 
relatively nearby pulsars, where there is less opportunity for underdensities and overdensities 
in the ISM along the line of sight to cancel, and hence it is plausible that distance models are 
on average more reliable for more distant pulsars.

\begin{figure}
\begin{center}
\includegraphics[width=0.9\textwidth]{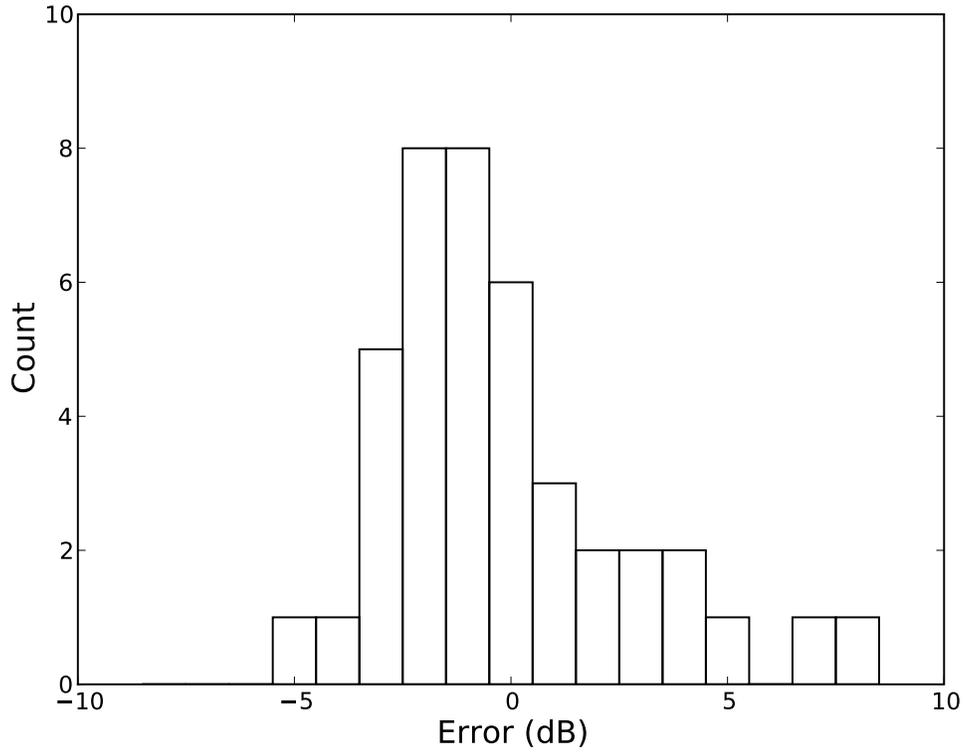}
\caption[Histogram of TC93 errors for pulsars with measured parallaxes]
{Histogram of TC93 errors for pulsars with measured parallaxes, with errors binned in 1 dB increments.
Underestimates of distance are more common in this model, but the largest errors are made when the
distance is overestimated.  The standard deviation of the errors is 2.8 dB.}
\label{fig:tc93hist}
\end{center}
\end{figure}

\begin{figure}
\begin{center}
\includegraphics[width=0.9\textwidth]{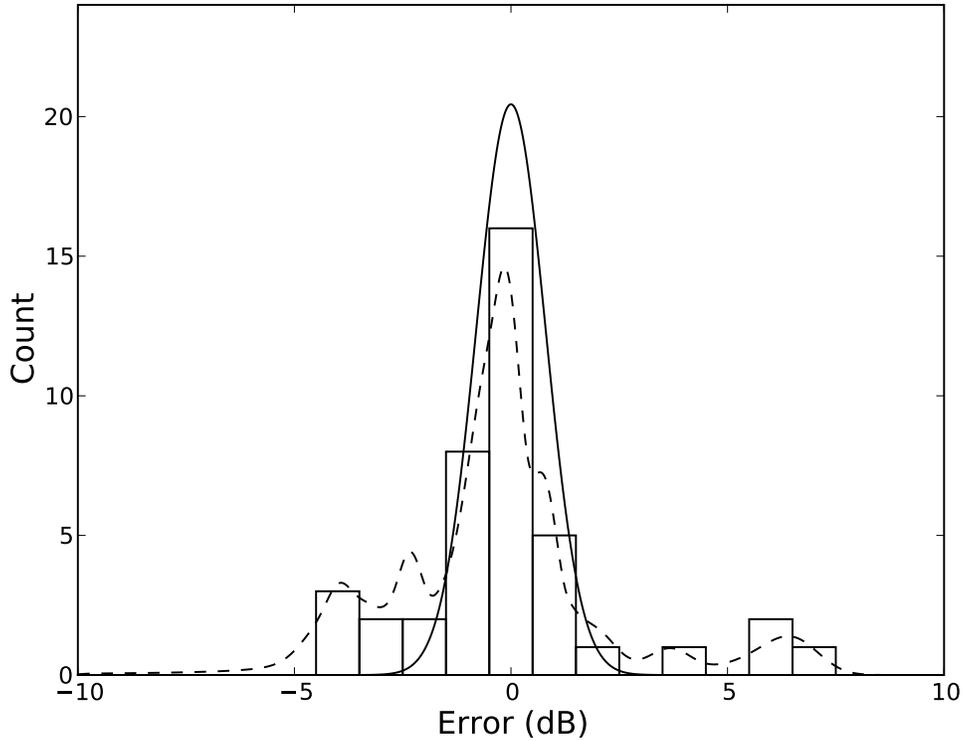}
\caption[Histogram of NE2001 errors for pulsars with measured parallaxes]
{Histogram of NE2001 errors for pulsars with measured parallaxes, with errors binned in 1 dB 
increments.  Unlike the TC93 model, there is no discernable systematic errors, but the distribution
is still clearly non--Gaussian, with a long tail of errors which can exceed 6 dB. As with TC93, the 
largest errors are seen when the distance is overestimated. 
The standard deviation of the errors is 2.3 dB, a small improvement on the 2.8 dB seen for TC93.
The ``binned" error model described in the text is shown as a dashed line, while
the single Gaussian ``traditional" model which approximates the 20\% errors commonly assumed 
for DM distances is shown as a solid line.}
\label{fig:ne2001hist}
\end{center}
\end{figure}

Figure~\ref{fig:tc93hist} shows a histogram of the observed distance errors for the TC93 model 
(where the best--fit parallax distances have been assumed to be the correct value), with 
the error expressed in dB, and binned in 1 dB increments.   The error in dB is defined as 
$10\log_{10}{\frac{\mathrm{model\ distance}}{\mathrm{parallax\ distance}}}$,
and hence positive values represent overestimates, and negative values represent underestimates.
Figure~\ref{fig:ne2001hist} repeats this plot for the NE2001 model.
Figure~\ref{fig:tc93hist} shows a clear systematic offset in the median error of the TC93 model 
towards underestimated distances, but the largest errors are seen when distances are overestimated.
Figure~\ref{fig:ne2001hist} shows that systematic bias has been largely removed by the NE2001 model,
but the distribution of errors still cannot be well approximated by a single Gaussian model,
due to the long wings towards large errors. 
For both the TC93 and NE2001 models, the largest errors are seen when 
the distance is overestimated.  The effect of these high--error ``outliers" on population studies
of neutron stars is considered further below.

Estimating the true form of the error distribution shown in Figure~\ref{fig:ne2001hist} is difficult
due to the small number of samples.  However, the single Gaussian model with standard
deviation 0.8 dB plotted as a solid line in Figure~\ref{fig:ne2001hist} shows that
the common assumption of 20\% errors for DM--based distance estimates is not realistic.  In the 
discussion below, this model (in which 67\% of errors are less than a factor of 1.2) is referred to
as the traditional error model. 

Rather than attempting to discern the true error distribution, we have modeled the distribution
by simply summing the probability distribution function of the distance error for each pulsar, which
accounts for the possible variation of the true distance from the parallax value.  The total probability
distribution function was binned in 0.001 dB increments and the result smoothed with a 
1 dB Hanning window.  This is plotted as a dashed line in Figure~\ref{fig:ne2001hist}, and is 
hereafter referred to as the binned error model. The inclusion of parallax uncertainties acts to broaden
the distribution slightly, and somewhat preferentially towards more negative errors.  This is caused by
a uniform parallax error encompassing a non--uniform distance space, with a greater range of
distances if the parallax was overestimated than underestimated (a manifestation of the so--called 
Lutz--Kelker bias; \citealt{lutz73a}).

\begin{figure}
\begin{center}
\includegraphics[width=0.9\textwidth]{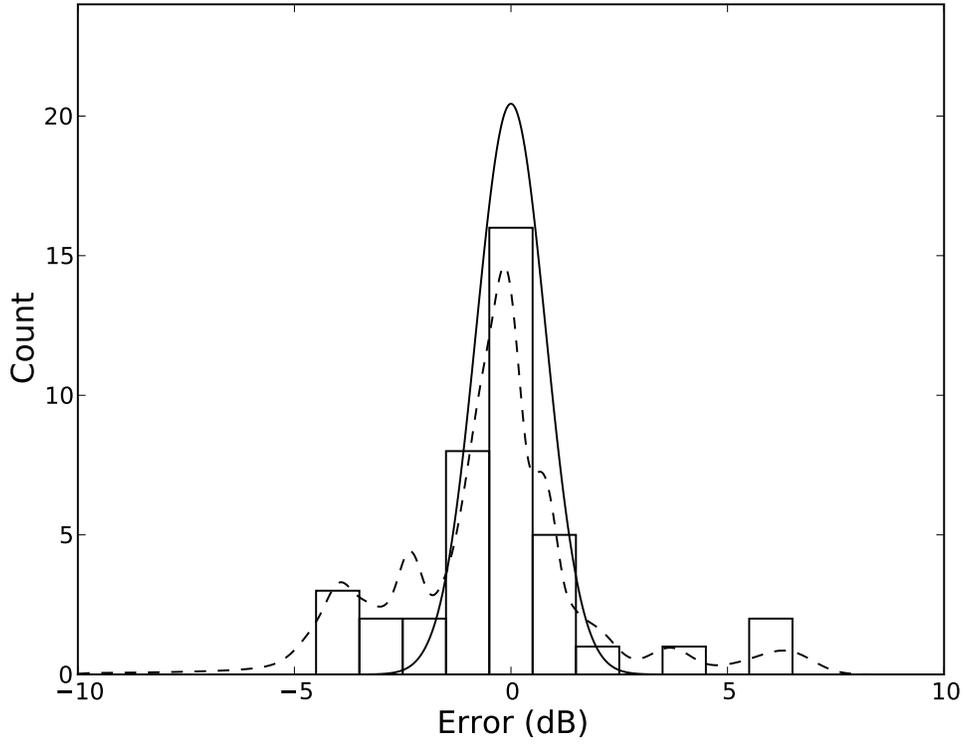}
\caption[Histogram of NE2001 errors for pulsars, excluding PSR B1541+09]
{Histogram of NE2001 errors for pulsars with measured parallaxes, with errors binned in 1 dB 
increments, after the exclusion of PSR B1541+09.  The ``binned" error model is shown as a 
dashed line, and the ``traditional" error model is shown as a solid line.  }
\label{fig:ne2001hist_no1541}
\end{center}
\end{figure}

Three pulsars with distance errors $>$5.5 dB dominate the tail of the NE2001 error distribution
with overestimated distances.  They are PSR J0613--0200 (NE2001 distance 1700 pc, parallax $2.1\pm0.6$ mas, 
\citealt{hotan06a}), \pthree\ (NE2001 distance 1450 pc, parallax $3.0\pm0.4$ mas, this work) and 
PSR B1541+09 (NE2001 distance 35000 pc, parallax $0.13\pm0.02$ mas, \citealt{chatterjee09a}).  
The NE2001 distance of 35 kpc for PSR B1541+09 is somewhat arbitrary, as the NE2001 model
does not possess sufficient electrons along the line of sight to PSR B1541+09 to account for its DM, and hence it is 
placed beyond the edge of the model.   This type of shortcoming is rare and can be easily identified, and so we
have excluded PSR B1541+09 from the analysis that follows.  Figure~\ref{fig:ne2001hist_no1541} shows the 
revised error histogram and binned model.  Despite the exclusion of PSR B1541+09, \pthree\ is still an example of 
the previously mentioned selection effect bias, and so (as already noted) large errors are likely to be somewhat 
over--represented in the subsequent analysis.  Thus, this model can safely be considered to be
a ``worst--case" representation of DM distance reliability.

\subsection{The pulsar velocity distribution}

Although is has been been shown that there is no evidence for systematic bias 
in the NE2001 distance model, errors exceeding 6 dB still exist for individual pulsars. Such 
errors can impact upon population analysis of pulsars, artificially creating tails of extremely high
or low values of distance--dependent parameters such as luminosity and velocity.  This can
affect estimates of the mean values of these quantities, as well
as confusing attempts to explain the underlying physics by generating false outliers.  As an example,
the case of neutron star transverse velocities is considered below.

\begin{figure}
\begin{center}
\includegraphics[width=0.9\textwidth]{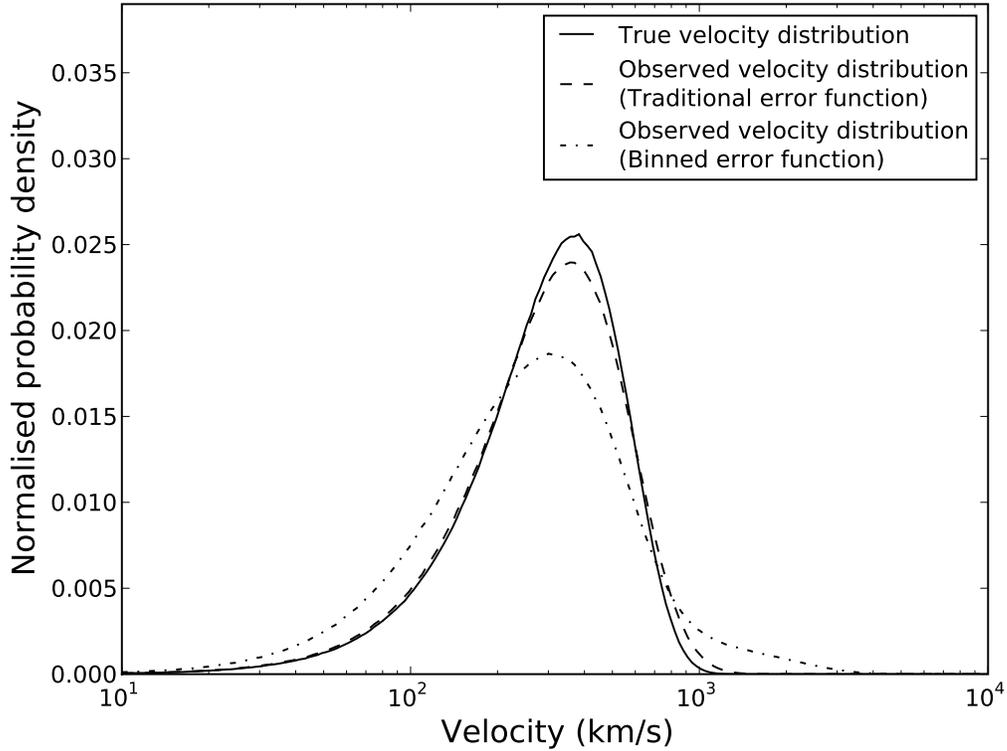}
\caption[Synthetic  2D velocity distribution]
{2D velocities for a synthetic pulsar population, as observed with no distance error (solid line), the
traditional distance error model (dashed) and the
binned distance error model described in the text (dash-dot).  The effect
of the binned distance error model is to broaden the observed velocity distribution, 
particularly at large velocities.}
\label{fig:disterror_velocity}
\end{center}
\end{figure}

\citet{hobbs05a} examine a large sample of pulsar proper motions and conclude that the distribution 
of 3D space velocities of young pulsars (characteristic age $< 3$\,Myr) is well fit by a Maxwellian
distribution with mean 431 \kms\ and one--dimensional rms 265 \kms.  This velocity distribution
is used as the starting point for the simulations that follow.  
A Monte Carlo simulation was performed, creating a synthetic pulsar catalogue of ten million
pulsars, where each pulsar's actual velocity was drawn from the 
distribution described above.  These are referred to as the unperturbed 
velocities.   No proper motion measurement error was assigned to the unperturbed velocities; more
distant pulsars (which possess on average a lower angular proper motion) would be more affected
by the presence of a roughly constant proper motion measurement error, which is another small
but unmodeled bias which would act to further broaden the observed velocity distribution.
The velocities of each pulsar were perturbed according to the binned and 
the traditional distance error functions described above, and the 
observed 2D velocity was recorded for each case.  
The observed 2D velocity distribution for the binned error function and the traditional error function,
along with the unperturbed 2D velocity distribution, is plotted in Figure~\ref{fig:disterror_velocity}.

It is immediately apparent that the broadening of the observed velocity distribution is
considerably greater for the binned error model, compared to the traditional error model, and  
is particularly pronounced at the high end of the velocity distribution.
The mean unperturbed 2D velocity is 332 \kms, which is increased by 2\% to 338\,\kms\ in
the traditional error model, and by 5\% to 349 \kms\ in the binned error model.

More important than the mean effect on pulsar velocities, however, is the effect on the high end 
of the pulsar velocity distribution.  \citet{hobbs05a} note that distance model inaccuracies could
be responsible for the sparse tail of very high velocity pulsars, but conclude that with the small
sample size available, the observed high--velocity pulsars are consistent with the tail of a 
continuous velocity distribution.  
In considering the impact of distance model errors, it is useful to examine the frequency
of occurrence of pulsars with observed transverse speeds greater than 1000\,\kms\ in the synthetic
catalogue.  These pulsars will be described as very high velocity (VHV) pulsars.  In the 
synthetic catalogue, only 0.1\% of pulsars fall into this category in the unperturbed velocity 
distribution, but the occurrence rises to 0.4\% when the traditional error model is applied, and
3.9\% for the binned error model.  By way of comparison, the ATNF pulsar catalogue contains
216 pulsars with listed transverse velocities -- of these, seven exceed 1000 \kms, approximately
3\% of the sample.

Thus, distance model errors can add to the high--velocity tail observed in the pulsar 
velocity distribution, an effect which is significantly enhanced when the distance errors ``outliers"
are accounted for.  These results suggest that the majority of VHV pulsars may in fact simply be 
pulsars with incorrect distance estimates. As already stated, the binned error model used here is likely
to overestimate the true frequency of large errors, and hence the true impact of DM distance
errors is likely to lie somewhere in between the binned error model and the traditional error model.  
Additionally, there is clear evidence that at least some VHV pulsars are present in the Galaxy, with
\citet{chatterjee05a} using VLBA astrometry to show that the transverse velocity of PSR B1508+55
exceeds 1000\,\kms, and observations of PSR B2224+65 showing scintillation, proper motion and
bow shock results consistent with a velocity close to 
1000 \kms\ \citep{cordes86b,harrison93a,cordes93a}. Nonetheless,
the binned error model is useful in illustrating that the largest impact of occasional, rare distance errors
is at the high end of the pulsar velocity distribution.
A larger, unbiased sample of pulsar
distance measurements would enable a more accurate estimation of the true distance error
function for NE2001.  This would in turn enable the effects of distance errors to be ``deconvolved"
from the measured pulsar velocity distribution with some confidence, and would lay the framework 
for a further improved Galactic electron density distribution model to supersede NE2001.

\section{Conclusion}
\label{sec:conclusions}
Using the Long Baseline Array, we have measured parallaxes for seven pulsars in the 
sparsely sampled southern sky, a four--fold increase on the previous number of southern 
VLBI parallaxes.  Based on the results of this survey, we conclude that parallaxes could be 
measured for at least 30 additional southern pulsars using the LBA in its present form, and 
future enhancement to LBA sensitivity can be expected to allow parallax measurements for
an increasing sample of fainter pulsars.
The ensemble of independent distance measures obtained from 
this program will allow significant improvements to Galactic electron models at these
latitudes.  Two of the measured distances (\pthree\ and \pfour) differed by more than a factor
of two from their DM estimates, reinforcing that DM--based distance estimates for individual pulsars
should be treated cautiously in the absence of any other distance indicators.  
Finally, it was shown that DM distance errors could be inflating 
estimates of both the mean pulsar velocity and the number of very high velocity pulsars, although the
current small and biased sample of independent distance measures makes this effect
difficult to quantify.

\acknowledgements

The authors would like to thank S. Chatterjee for providing the fitted positions of \psix\ from 
VLBA observations.
The Long Baseline Array is part of the Australia Telescope which is funded by the Commonwealth of Australia for operation as a National Facility managed by CSIRO.  This research made use of the
ATNF Pulsar Catalogue and the NASA/IPAC Extragalactic Database (NED). NED is operated by 
the Jet Propulsion Laboratory, California Institute of Technology, under contract with the 
National Aeronautics and Space Administration.

\bibliographystyle{apj}
\bibliography{deller_thesis}

\clearpage

\begin{deluxetable}{lccccccc}
\tabletypesize{\tiny}
\tablecaption{Target pulsars}
\tablewidth{0pt}
\tablehead{
\colhead{Pulsar} & \colhead{TC93 distance} & \colhead{NE2001 distance} & \colhead{1600 MHz} & \colhead{Pulsar} & Equivalent gated & \colhead{Reference} & \colhead{Calibrator/target} \\
\colhead{name} & \colhead{(pc)} & \colhead{(pc)} & \colhead{flux (mJy)} & \colhead{gating gain} & 1600 MHz flux (mJy) & \colhead{source} & \colhead{separation (deg)} 
}
\startdata
PSR J0108--1431 		& 130 	& 180	& 0.8\tablenotemark{a} 	& 5.1 	& 4		& J0111--1317 	& 1.5 \\
PSR J0437--4715 		& 140 	& 190	& 140 				& 6.25	& 875	& J0439--4522 	& 1.9 \\
PSR J0630--2834 		& 2150 	& 1450	& 23 					& 3.5 	& 81		& J0628--2805 	& 0.7 \\
PSR J0737--3039A/B 	& 570 	& 570	& 1.6 				& 2.5 	& 4		& J0738--3025 	& 0.4 \\
PSR J1559--4438 		& 1600 	& 2350	& 40 					& 3.6 	& 144	& J1604--4441 	& 0.9 \\
PSR J2048--1616 		& 640 	& 560	& 13 					& 3.6		& 47		& J2047--1639 	& 0.5 \\
PSR J2144--3933 		& 180 	& 264	& 0.8\tablenotemark{a} 	& 10 		& 8		& J2141--3729 	& 2.1 \\
PSR J2145--0750 		& 500 	& 570	& 8\tablenotemark{a} 	& 4.3 	& 34		& J2142--0437 	& 3.3 \\
\enddata
\tablenotetext{a}{Pulsar suffers heavily from long timescale scintillation, so individual epochs vary considerably from the average value shown.}
\label{tab:pulsars}
\end{deluxetable}

\begin{deluxetable}{llll}
\tabletypesize{\tiny}
\tablecaption{Reference source positions}
\tablewidth{0pt}
\tablehead{
\colhead{Source} & \colhead{Right ascension (J2000)} & \colhead{Declination (J2000)} & \colhead{Reference}
}
\startdata
J0111--1317 	&  01$^{\mathrm h}$11$^{\mathrm m}$56.857962$^{\mathrm s}$
			&  $-$13\degrees17'01.19702"
			& VCS5; \citet{kovalev07a} \\
J0439--4522 	&  04$^{\mathrm h}$39$^{\mathrm m}$00.854714$^{\mathrm s}$
			&  $-$49\degrees22'22.56260"
			& ICRF; \citet{ma98a} \\
J0628--2805 	&  06$^{\mathrm h}$28$^{\mathrm m}$43.279032$^{\mathrm s}$
			&  $-$28\degrees05'19.38364"
			& VCS5; \citet{kovalev07a} \\
B0736--303 	&  07$^{\mathrm h}$38$^{\mathrm m}$19.78785$^{\mathrm s}$
			&  $-$30\degrees25'04.8360"
			& None; see \citet{deller09a} \\
J1604--4441 	&  16$^{\mathrm h}$04$^{\mathrm m}$31.020692$^{\mathrm s}$
			&  $-$44\degrees41'31.97421"
			& ICRF; \citet{ma98a} \\
J2047--1639 	&  20$^{\mathrm h}$47$^{\mathrm m}$19.667018$^{\mathrm s}$
			&  $-$16\degrees39'05.84249"
			& VCS1; \citet{beasley02a} \\
J2141--3729 	&  21$^{\mathrm h}$41$^{\mathrm m}$52.448964$^{\mathrm s}$
			&  $-$37\degrees29'12.99259"
			& VCS3; \citet{petrov05a} \\
J2142--0437 	&  21$^{\mathrm h}$42$^{\mathrm m}$36.901688$^{\mathrm s}$
			&  $-$04\degrees37'43.51275"
			& VCS1; \citet{beasley02a} \\
\enddata
\label{tab:calpos}
\end{deluxetable}

\clearpage
\input{allresults.tab}
\clearpage

\end{document}

%% file: allresults.tab.tex
\begin{deluxetable}{lrrrr}
\tabletypesize{\tiny}
\tablecaption{Astrometric fits for all target pulsars}
\tablewidth{0pt}
\rotate
\tablehead{
\colhead{Parameter} & \colhead{PSR J0108-1431} & \colhead{PSR J0437--4715} & 
\colhead{PSR J0630--2834} & \colhead{PSR J0737--3039}
}
\startdata
Right ascension (J2000)         & 01:08:08.347016 $\pm$ 0.000088
                                                        & 04:37:15.883250 $\pm$ 0.000003
                                                        & 06:30:49.404393 $\pm$ 0.000043
                                                & 07:37:51.248419 $\pm$ 0.000026               \\[0.5ex]
Declination (J2000)             & -14:31:50.187139 $\pm$ 0.001069
                                        & -47:15:09.031863 $\pm$ 0.000037
                                                & -28:34:42.778813 $\pm$ 0.000372
                                        & -30:39:40.714310 $\pm$ 0.000099               \\[0.5ex]
Proper motion in right ascension ($\mu_{\alpha}$; mas yr$^{-1}$)         & 75.05 $\pm$ 2.26\phn\phn\phn\phn
                                                        & 121.679 $\pm$ 0.05\phn\phn\phn\phn
                                                        & -46.30 $\pm$ 0.99\phn\phn\phn\phn
                                                        & -3.82 $\pm$ 0.62\phn\phn\phn\phn      \\[0.5ex]
Proper motion in declination ($\mu_{\delta}$; mas yr$^{-1}$)         & -152.54 $\pm$ 1.65\phn\phn\phn\phn
                                                        & -71.820 $\pm$ 0.09\phn\phn\phn\phn
                                                        & 21.26 $\pm$ 0.52\phn\phn\phn\phn
                                                        & 2.13 $\pm$ 0.23\phn\phn\phn\phn       \\[0.5ex]
Parallax ($\pi$; mas)     & 4.170 $\pm$ 1.421\phn\phn\phn
                        & 6.396 $\pm$ 0.054\phn\phn\phn
                        & 3.009 $\pm$ 0.409\phn\phn\phn
                        & 0.872 $\pm$ 0.143\phn\phn\phn \\[0.5ex]
Distance (pc)   & $240^{+124}_{-61}$\phs \phn\phd\phn\phn\phn\phn\phn\phn$\  $
                        & $156.3^{+1.3}_{-1.3}$\phs \phn\phd\phn\phn\phn\phn\phn\phn$\  $
                        & $332^{+52}_{-40}$\phs \phn\phd\phn\phn\phn\phn\phn\phn$\  $
                        & $1150^{+220}_{-160}$\phs \phn\phd\phn\phn\phn\phn\phn\phn$\  $        \\[0.5ex]
Transverse velocity ($v_{t}$; km s$^{-1}$)   & $194^{+104}_{-51}$\phs \phn\phd\phn\phn\phn\phn\phn\phn$\  $
                                        & $104.7^{+1.0}_{-1.0}$\phs \phn\phd\phn\phn\phn\phn\phn\phn$\  $
                                        & $80^{+15}_{-11}$\phs \phn\phd\phn\phn\phn\phn\phn\phn$\  $
                                        & $24^{+9}_{-6}$\phs \phn\phd\phn\phn\phn\phn\phn\phn$\  $      \\[0.5ex]
Reference epoch (MJD) & 54100.0\phs \phn\phd\phn\phn\phn\phn\phn\phn$\  $
                                            & 54100.0\phs \phn\phd\phn\phn\phn\phn\phn\phn$\  $
                                            & 54100.0\phs \phn\phd\phn\phn\phn\phn\phn\phn$\  $
                                            & 54100.0\phs \phn\phd\phn\phn\phn\phn\phn\phn$\  $ \\[0.5ex]
Visibility weighting            & Sensitivity\phs \phn\phd\phn\phn\phn\phn\phn\phn$\  $
                                        & Equal\phs \phn\phd\phn\phn\phn\phn\phn\phn$\  $
                                        & Equal\phs \phn\phd\phn\phn\phn\phn\phn\phn$\  $
                                        & Sensitivity\phs \phn\phd\phn\phn\phn\phn\phn\phn$\  $ \\[0.5ex]
Image weighting         & Natural\phs \phn\phd\phn\phn\phn\phn\phn\phn$\  $
                                        & Uniform\phs \phn\phd\phn\phn\phn\phn\phn\phn$\  $
                                        & Uniform\phs \phn\phd\phn\phn\phn\phn\phn\phn$\  $
                                        & Natural\phs \phn\phd\phn\phn\phn\phn\phn\phn$\  $             \\[0.5ex]
Average epoch mean fit error (mas)      & 1.232\phs \phn\phd\phn\phn\phn\phn\phn\phn$\  $
                                                        & 0.059\phs \phn\phd\phn\phn\phn\phn\phn\phn$\  $
                                                        & 0.765\phs \phn\phd\phn\phn\phn\phn\phn\phn$\  $
                                                        & 0.747\phs \phn\phd\phn\phn\phn\phn\phn\phn$\  $       \\[0.5ex]
Average intra--epoch systematic error (mas)     & 2.477\phs \phn\phd\phn\phn\phn\phn\phn\phn$\  $
                                                                        & 0.068\phs \phn\phd\phn\phn\phn\phn\phn\phn$\  $
                                                                        & 0.839\phs \phn\phd\phn\phn\phn\phn\phn\phn$\  $
                                                                        & 0.939\phs \phn\phd\phn\phn\phn\phn\phn\phn$\  $ 
                                                                        \\[0.5ex]
Average inter--epoch systematic error (mas)     & 4.310\phs \phn\phd\phn\phn\phn\phn\phn\phn$\  $ 
                                                                        & 0.103\phs \phn\phd\phn\phn\phn\phn\phn\phn$\  $ 
                                                                        & 1.205\phs \phn\phd\phn\phn\phn\phn\phn\phn$\  $
                                                                        & 0.0\phs \phn\phd\phn\phn\phn\phn\phn\phn$\  $ 
                                                                        \\[0.5ex]
Average single--epoch SNR	& 8\phs \phn\phd\phn\phn\phn\phn\phn\phn$\  $
                                                & 21\phs \phn\phd\phn\phn\phn\phn\phn\phn$\  $
                                                & 15\phs \phn\phd\phn\phn\phn\phn\phn\phn$\  $
                                                & 17\phs \phn\phd\phn\phn\phn\phn\phn\phn$\  $  \\[0.5ex]
Covariance of $\pi$ and $\mu_{\alpha}$  & 0.27\phs \phn\phd\phn\phn\phn\phn\phn\phn$\  $
                                                & -0.39\phs \phn\phd\phn\phn\phn\phn\phn\phn$\  $
                                                & -0.01\phs \phn\phd\phn\phn\phn\phn\phn\phn$\  $
                                                & -0.01\phs \phn\phd\phn\phn\phn\phn\phn\phn$\  $  \\[0.5ex]
Covariance of $\pi$ and $\mu_{\delta}$  & 0.00\phs \phn\phd\phn\phn\phn\phn\phn\phn$\  $
                                                & 0.60\phs \phn\phd\phn\phn\phn\phn\phn\phn$\  $
                                                & -0.10\phs \phn\phd\phn\phn\phn\phn\phn\phn$\  $
                                                & -0.06\phs \phn\phd\phn\phn\phn\phn\phn\phn$\  $  \\[1.0ex]
\hline\hline
&&&&\\[-4pt]
\multicolumn{1}{c}{\tiny Parameter} & 
\multicolumn{1}{c}{\tiny PSR J1559--4438} & 
\multicolumn{1}{c}{\tiny PSR J2048--1616} & 
\multicolumn{1}{c}{\tiny PSR J2144--3933} & 
\multicolumn{1}{c}{\tiny PSR J2145--0750\tablenotemark{A}} \\[1.0ex]
\hline
Right ascension (J2000)         & 15:59:41.526126 $\pm$ 0.000008
                                                & 20:48:35.640637 $\pm$ 0.000040
                                                & 12:44:12.060404 $\pm$ 0.000045
                                                & 21:45:50.461901 $\pm$ 0.000098                \\[0.5ex]
Declination (J2000)             & -44:38:45.901778 $\pm$ 0.000035
                                        & -16:16:44.553501 $\pm$ 0.000147
                                        & -39:33:56.885041 $\pm$ 0.000316
                                        & -07:50:18.462388 $\pm$ 0.000558                \\[0.5ex]
Proper motion in right ascension ($\mu_{\alpha}$; mas yr$^{-1}$)         & 1.52 $\pm$ 0.14\phn\phn\phn\phn
                                                        & 114.24 $\pm$ 0.52\phn\phn\phn\phn
                                                        & -57.89 $\pm$ 0.88\phn\phn\phn\phn
                                                        & -15.43 $\pm$ 2.07\phn\phn\phn\phn     \\[0.5ex]
Proper motion in declination ($\mu_{\delta}$; mas yr$^{-1}$)         & 13.15 $\pm$ 0.05\phn\phn\phn\phn
                                                        & -4.03 $\pm$ 0.24\phn\phn\phn\phn
                                                        & -155.90 $\pm$ 0.54\phn\phn\phn\phn
                                                        & -7.67 $\pm$ 0.81\phn\phn\phn\phn      \\[0.5ex]
Parallax ($\pi$; mas)     & 0.384 $\pm$ 0.081\phn\phn\phn
                        & 1.712 $\pm$ 0.909\phn\phn\phn
                        & 6.051 $\pm$ 0.560\phn\phn\phn
                        & --\phs \phn\phd\phn\phn\phn\phn\phn\phn$\  $          \\[0.5ex]
Distance (pc)   & $2600^{+690}_{-450}$\phs \phn\phd\phn\phn\phn\phn\phn\phn$\  $
                        & $580^{+660}_{-200}$\phs \phn\phd\phn\phn\phn\phn\phn\phn$\  $
                        & $165^{+17}_{-14}$\phs \phn\phd\phn\phn\phn\phn\phn\phn$\  $
                        & --\phs \phn\phd\phn\phn\phn\phn\phn\phn$\  $          \\[0.5ex]
Transverse velocity ($v_{t}$; km s$^{-1}$)   & $163^{+44}_{-29}$\phs \phn\phd\phn\phn\phn\phn\phn\phn$\  $
                                        & $317^{+362}_{-111}$\phs \phn\phd\phn\phn\phn\phn\phn\phn$\  $
                                        & $130^{+14}_{-12}$\phs \phn\phd\phn\phn\phn\phn\phn\phn$\  $
                                        & --\phs \phn\phd\phn\phn\phn\phn\phn\phn$\  $          \\[0.5ex]
Reference epoch (MJD) & 54100.0\phs \phn\phd\phn\phn\phn\phn\phn\phn$\  $
                                            & 54100.0\phs \phn\phd\phn\phn\phn\phn\phn\phn$\  $
                                            & 54100.0\phs \phn\phd\phn\phn\phn\phn\phn\phn$\  $
                                            & 54100.0\phs \phn\phd\phn\phn\phn\phn\phn\phn$\  $ \\[0.5ex]
Visibility weighting            & Equal\phs \phn\phd\phn\phn\phn\phn\phn\phn$\  $
                                        & Equal\phs \phn\phd\phn\phn\phn\phn\phn\phn$\  $
                                        & Equal\phs \phn\phd\phn\phn\phn\phn\phn\phn$\  $
                                        & Sensitivity\phs \phn\phd\phn\phn\phn\phn\phn\phn$\  $         \\[0.5ex]
Image weighting         & Uniform\phs \phn\phd\phn\phn\phn\phn\phn\phn$\  $
                                        & Uniform\phs \phn\phd\phn\phn\phn\phn\phn\phn$\  $
                                        & Natural\phs \phn\phd\phn\phn\phn\phn\phn\phn$\  $
                                        & Natural\phs \phn\phd\phn\phn\phn\phn\phn\phn$\  $             \\[0.5ex]
Average epoch mean fit error (mas)      & 0.242\phs \phn\phd\phn\phn\phn\phn\phn\phn$\  $
                                                        & 0.517\phs \phn\phd\phn\phn\phn\phn\phn\phn$\  $
                                                        & 1.025\phs \phn\phd\phn\phn\phn\phn\phn\phn$\  $
                                                        & 2.136\phs \phn\phd\phn\phn\phn\phn\phn\phn$\  $       \\[0.5ex]
Average intra--epoch systematic error (mas)     & 0.259\phs \phn\phd\phn\phn\phn\phn\phn\phn$\  $
                                                                        & 1.282\phs \phn\phd\phn\phn\phn\phn\phn\phn$\  $
                                                                        & 1.450\phs \phn\phd\phn\phn\phn\phn\phn\phn$\  $
                                                                        & 0.0\phs \phn\phd\phn\phn\phn\phn\phn\phn$\  $ 
                                                                        \\[0.5ex]
Average inter--epoch systematic error (mas)     & 0.055\phs \phn\phd\phn\phn\phn\phn\phn\phn$\  $
                                                                        & 0.105\phs \phn\phd\phn\phn\phn\phn\phn\phn$\  $
                                                                        & 0.875\phs \phn\phd\phn\phn\phn\phn\phn\phn$\  $
                                                                        & 0.0\phs \phn\phd\phn\phn\phn\phn\phn\phn$\  $ 
                                                                        \\[0.5ex]
Average single--epoch SNR       & 50\phs \phn\phd\phn\phn\phn\phn\phn\phn$\  $
                                                & 23\phs \phn\phd\phn\phn\phn\phn\phn\phn$\  $
                                                & 10\phs \phn\phd\phn\phn\phn\phn\phn\phn$\  $
                                                & 8\phs \phn\phd\phn\phn\phn\phn\phn\phn$\  $  \\[0.5ex]
Covariance of $\pi$ and $\mu_{\alpha}$  & -0.43\phs \phn\phd\phn\phn\phn\phn\phn\phn$\  $
                                                & 0.78\phs \phn\phd\phn\phn\phn\phn\phn\phn$\  $
                                                & 0.23\phs \phn\phd\phn\phn\phn\phn\phn\phn$\  $
                                                & --\phs \phn\phd\phn\phn\phn\phn\phn\phn$\  $  \\[0.5ex]
Covariance of $\pi$ and $\mu_{\delta}$  & 0.18\phs \phn\phd\phn\phn\phn\phn\phn\phn$\  $
                                                & 0.64\phs \phn\phd\phn\phn\phn\phn\phn\phn$\  $
                                                & 0.16\phs \phn\phd\phn\phn\phn\phn\phn\phn$\  $
                                                & --\phs \phn\phd\phn\phn\phn\phn\phn\phn$\  $   \\[0.5ex]
\enddata
\tablenotetext{A}{Based on two detections, with parallax fixed at 0.  Variation of the parallax value 
between 0 and 2 mas results in less than 100 $\mu$as yr$^{-1}$\ difference in derived proper motion.}
\label{tab:allresults}
\end{deluxetable}